\documentclass[preprint,showkeys,preprintnumbers,amsmath,amssymb]{revtex4}

\usepackage{graphicx}
\usepackage{dcolumn}
\usepackage{bm}
\usepackage{subfigure}
\usepackage{float}
\usepackage{lineno}

\begin{document}
\title{\textbf{\textrm{Beta-rhythm oscillations and synchronization transition in network models of Izhikevich neurons: effect of topology and synaptic type}}}

\author{Mahsa Khoshkhou}
\author{Afshin Montakhab}
 \email{montakhab@shirazu.ac.ir}

\affiliation{Department of Physics, College of Sciences, Shiraz University, Shiraz 71946-84795, Iran}

\date{\today}
\begin{abstract}
Despite their significant functional roles, beta-band oscillations
are least understood. Synchronization in neuronal networks have
attracted much attention in recent years with the main focus on
transition type. Whether one obtains explosive transition or a
continuous transition is an important feature of the neuronal
network which can depend on network structure as well as synaptic
types. In this study we consider the effect of synaptic
interaction (electrical and chemical) as well as structural
connectivity on synchronization transition in network models of
Izhikevich neurons which spike regularly with beta rhythms. We
find a wide range of behavior including continuous transition,
explosive transition, as well as lack of global order. The
stronger electrical synapses are more conducive to synchronization
and can even lead to explosive synchronization. The key network
element which determines the order of transition is found to be
the clustering coefficient and not the small world effect, or the
existence of hubs in a network. These results are in contrast to
previous results which use phase oscillator models such as the
Kuramoto model. Furthermore, we show that the patterns of
synchronization changes when one goes to the gamma band. We
attribute such a change to the change in the refractory period of
Izhikevich neurons which changes significantly with frequency.
\end{abstract}


\maketitle
\section{introduction}
Synchronization is an important collective phenomenon that may
emerge in locally interacting physical and biological oscillatory
systems \cite{Pikovsky,Arenas,Motter,Barahona}. Neural tissue of
central nervous system can generate oscillatory activity in
various scales from individual neuron firing to macroscopic
oscillations in large neural ensembles
\cite{Buzsaki,Varela,Engel2001,Buzsaki2004,Jensen2005}.
Macroscopic rhythmic activity which is observed in
electroencephalography (EEG) recordings, is believed to occur due
to emergence of synchronization in oscillations of constituent
neurons. Synchronization of neural activity has a fundamental role
in brain functions such as vision, memory, action, perception,
information transfer, thought and so on
\cite{Varela,Kandel,Steinmetz,Fries,Fell,Uhlhass,Jiang}.

Neural oscillations have been documented to cover a broad spectrum
of frequencies. These oscillations are observed widely in every
level of central nervous system and are usually categorized into
five frequency bands: delta $0.5-3.5$ Hz, theta $4-7$ Hz, alpha
$8-12$ Hz, beta $13-30$ Hz and gamma $>30$ Hz
\cite{Buzsaki,Rosanova2009}. Beta rhythms are associated with
normal wakeful consciousness states and appear when one is alert,
attentive or when a person is engaged in problem solving or
decision making. Beta waves are also associated with the
activities of motor cortex \cite{Engel2010,Baker}.

Synchronization in neural population has been in the focus of
intense experimental and theoretical research recently. See
\cite{Bennett,Mikkelsen,Valizadeh2016,Pazo,Zhou,Matias,Valizadeh2014,Salinas}
for a few examples. Although beta-band activities have a
significant role in brain functions, they have attracted less
attention than other frequency bands \cite{Engel2010}. This is all
the more important as many fundamental functions of the brain are
associated with such oscillations. For example, synchronization
transition is an important issue.  From a theoretical point of
view, synchronization in a neuronal network occurs as one
increases synaptic strength.  How this transition occurs is of
fundamental importance. Generally, the transition can occur either
as a continuous transition or a discontinuous (explosive) manner.
If continuous, a small change can lead to small changes in systems
response; however, if explosive, a small change can lead to
dramatic changes in system's response. In addition to the type of
synaptic interaction, the role of network topology is of key issue
in determining the order of synchronization transition. In this
paper, we intend to investigate the effect of network topology and
synaptic type on synchronization phase transition in populations
of spiking neurons with none-identical intrinsic frequencies in
beta band. Specifically, we will focus on the order of the
emerging phase transition for various network structures and
different synaptic interactions. It is believed that normal brain
activity requires it to be close to a phase boundary (a critical
point) which consequently provide access to both synchronous and
asynchronous oscillations with small change in the input
\cite{Beggs2012,Hesse2014,MMV2017}. Hence, it is important to know
whether the emerging synchronization transition is continuous or
abrupt.

It is usual to evoke phase oscillators to characterize transition
properties of neural oscillations. See
\cite{Kitzbichler,Cumin,Timms,Shimokawa,Botcharova,Maistrenko,Gardenes2010}
for some examples. While this choice offers many computational and
analytic advantages, it suffers from some drawbacks.  For example,
it is not possible to consider a biologically realistic dynamical
model as a phase oscillator, since many important features such as
realistic synaptic interaction, are not easily implemented in
phase oscillator models such as the Kuramoto model.  Also, the
spiking patterns of real neurons with wide range of frequencies
are washed out in phase oscillator models. We therefore propose to
study neuronal dynamics according to the Izhikevich model
\cite{Izhikevich2003} which is obtained by reducing some
biological aspects of Hodgkin-Huxley (HH) neuron using bifurcation
methods \cite{Izhikevich2007}. This model is computationally
simpler than HH neuron, but is still biologically plausible.

To describe the functional form of synaptic interactions, we use
two experimentally documented synaptic types: electrical synapses
or gap junctions and chemical synapses \cite{Roth,
Perez,Kuo2016,Kopell2004,Simon2005,Sohal2005}. These two types of
interactions will appear as distinct expressions for synaptic
currents to be added to the Izhikevich neurons. To describe the
structure of synaptic interaction, we couple neurons via a
network. It is well-known that network connectivity can have
strong effects on patterns of collective behavior such as
synchronization \cite{Watts,Strogatz}. It is believed that key
elements such as small-world effect, clustering, and heterogeneity
are of fundamental importance effecting the general collective
behavior of a network. We therefore propose to study various
network structures starting with a regular ring with high
clustering and no randomness. We next consider small-world
networks which provide a balance between high clustering and
small-world effect.  We also consider the more random structures
such as Erdos-Renyi (homogeneous) and scale-free (heterogeneous)
networks with low clustering but dense long-range synapses.

Our main results are as follows: (i) we find that electrical
synapses are more conducive to synchronization than chemical
synapses, leading to explosive synchronization in beta band in
random networks. (ii) we find that the effect of clustering is far
more important than small-world effect in determining the order of
transition. (iii) we find that patterns of synchronization are
distinctly different in beta band from the corresponding
transitions in the high frequency gamma band.

\begin{table}
\center \caption{\small Theoretical values of clustering
coefficient $C$ and its corresponding measured values for the four
network structures we have used in this study, first and second
columns. Similar results for average path length $L$ are also
shown in the third and fourth columns. Note that the theoretical
values for WS network depend on the density of long-range links
$p$ and is not available in the given closed form, see Fig.3(a)
for more details. Theoretical calculations are performed by using
formulas in \cite{Gros}. The size of all networks is $N=1000$ and
the coordination number is $z{\simeq}50$ except for SF network
where $z{\simeq}20$.} \label{t1}
\begin{center}
\begin{tabular}{|c|c|c|c|c|}
\hline
$$ Network & $C_t$ & $C_m$ & $L_t$ & $L_m$ \\
\hline
Ring & 0.734 & 0.734 & 10.5 & 10.5 \\
\hline
WS & $-$ & 0.730 & $-$ & 3.83 \\
\hline
ER & 0.048 & 0.054 & 1.77 & 2.06 \\
\hline
SF & 0.020 & 0.026 & 2.85 & 2.60 \\
\hline
\end{tabular}
\end{center}

\end{table}

\section{methods}
To construct a neural circuit, we consider $N$ Izhikevich neurons
on an arbitrary network with a specific  (symmetric) adjacency
matrix $A$. The electrical activity of each neuron of this
ensemble is described by a set of two ordinary nonlinear coupled
differential equations \cite{Izhikevich2003}:
  \begin{equation}\label{equ1}
\frac{dv_i}{dt}=0.04v_i^2+5v_i+140-u_i+I_i^{DC}+I_i^{syn}
\end{equation}
\begin{equation}\label{equ2}
\frac{du_i}{dt}=a(bv_i-u_i)
\end{equation}
\\with the auxiliary after-spike reset:
\begin{equation}\label{equ3}
if\ \ v_i{\geq}30,\ \ then \ v_i \ {\rightarrow} \ c \ \ and\ u_i \ {\rightarrow} \ u_i+d
\end{equation}
\\
for $i=1, 2,..., N$. Here $v_i$ is the membrane potential and
$u_i$ is the membrane recovery variable. When $v_i$ reaches its
apex ($v_{max}=30$ mV), voltage and recovery variable are reset
according to Eq.(3). The term $(0.04v_i^2+5v_i+140)$ has been
chosen so that $v$ has \text{mV} unit and $t$ has \text{ms} units
\cite{Izhikevich2003}. In addition $a$, $b$, $c$ and $d$ are four
adjustable parameters in this model. Tuning these parameters,
Izhikevich neuron is capable of reproducing about twenty different
intrinsic firing patterns observed in real neurons
\cite{Izhikevich2006,Izhikevich2008}. In this paper we set
$a=0.02$, $b=0.2$, $c=-65$ and $d=8$, which corresponds to regular
spiking pattern \cite{Izhikevich2003,Izhikevich2006}.

The term $I_i^{DC}$ is an external current which determines
intrinsic firing rate of uncoupled Izhikevich neurons. Regularly
spiking Izhikevich neurons exhibits a Hopf bifurcation at
$I^{DC}=3.78$ \cite{Kim}. We choose values of $I_i^{DC}$ randomly
from a Poisson distribution with mean value $10$. Thus the
intrinsic firing rates $f_i$ lay in beta band and are different
from one neuron to the other. The term $I_i^{syn}$ in Eq.(1)
denotes synaptic current received by post-synaptic neuron $i$. If
the synapse is electrical, the synaptic current is
\cite{Roth,Perez}:
\begin{equation}\label{equ4}
    I_i^{syn}=\frac{1}{D_i}\sum_jg_{ji}(v_j-v_i)
\end{equation}
\\and if the synapse is chemical then \cite{Roth,Perez}:
\begin{equation}\label{equ5}
I_i^{syn}=\frac{1}{D_i}\sum_jg_{ji}\frac{exp(-\frac{t-t_j}{\tau_s})-exp(-\frac{t-t_j}{\tau_f})}{\tau_s-\tau_f}(V_0-v_i)
\end{equation}
\\
where $D_i$ is in-degree of node $i$, $g_{ji}$ is the strength of
synapse from pre-synaptic neuron $j$ to post-synaptic neuron $i$.
$g_{ji}=g A_{ji}$, where $g$ is the electrical conductance of
synapse and $A_{ji}$ is the element of adjacency matrix of the
underlying network \cite{Gros}. $A_{ji}=1$ if nodes $j$ and $i$
are connected and $A_{ji}=0$, otherwise. Also in Eq.(5)
$\tau_s=1.7$ and $\tau_f=0.2$ are the slow and fast synaptic decay
constants \cite{Roth}, $t_j$ is the instance of last spike of
pre-synaptic neuron $j$ and $V_0$ is the reversal potential of
synapse which is equal to zero since we assumed that all synapses
in our circuit are excitatory. We only consider networks which are
composed of one giant cluster, and thus no isolated nodes or
clusters exist.

Our main goal is to study the role of various network properties
on synchronization patterns that may emerge.  The leading network
properties we consider are clustering coefficient ($C$) and
average path length ($L$).  Also, the existence of hubs in
heterogeneous networks are thought to play an important role in
synchronization.  We consider a regular ring with high clustering
and large average path length (large-world-effect), a slightly
random network which preserves clustering but has small-world
effect, as well as two random networks which exhibit small
clustering but strong small-world effect: a homogeneous Poissonian
network with no hub as well as a heterogeneous scale-free network
with hubs. Details of the networks used and comparison with
corresponding theoretical values are summarized in Table I.

We integrate the dynamical equations using fourth-order
Runge-Kutta method with a time step of $0.01$ \text{ms} in order
to obtain $v_i(t)$. We typically evolve the entire system for a
long time and make sure that the system has reached a stationary
state. We then perform our measurements and calculations. We
obtain the instants of firings of all neurons and then assign an
instantaneous phase to each neuron between each pairs of
successive spikes, as in \cite{Pikovsky1997}:
    \begin{equation}\label{equ6}
        \phi_i(t)=2\pi\frac{t-t_i^m}{t_i^{m+1}-t_i^m}
    \end{equation}
\\where $t_i^m$ is the instant of $m^{th}$ spike of neuron $i$. We define a global instantaneous order parameter:
    \begin{equation}\label{equ7}
        S(t)=\frac{2}{N(N-1)}\sum_{i\neq j}cos^2\Big{(}\frac{\phi_i(t)-\phi_j(t)}{2}\Big{)}
    \end{equation}
\\
where the sum is over all pairs of neurons in the system whether
they are connected or not. The global order parameter $S$ is the
long-time-average of $S(t)$ in the stationary state of the system
($S={\langle}S(t){\rangle}_t$) and measures the collective phase
synchronization in neuronal oscillations. $S$ is bounded between
0.5 and 1. If neurons spike out-of-phase, then $S{\simeq}0.5$,
where they spike completely in-phase $S{\simeq}1$ and for states
with partial synchrony $0.5<S<1$. Synchronization transition is
displayed in $S-g$ plots where transition is expected to occur at
a given value of $g$. We note that we have also calculated the
Kuramoto order parameter and have found identical results as the
ones calculated using Eq.(\ref{equ7}). The relevant codes have
been shared in public domain at \texttt{figshare.com}.

\section{results}
\begin{figure}[!htbp]
\begin{center}
\subfigure[]{\includegraphics[width=0.53\textwidth,height=0.25\textwidth]{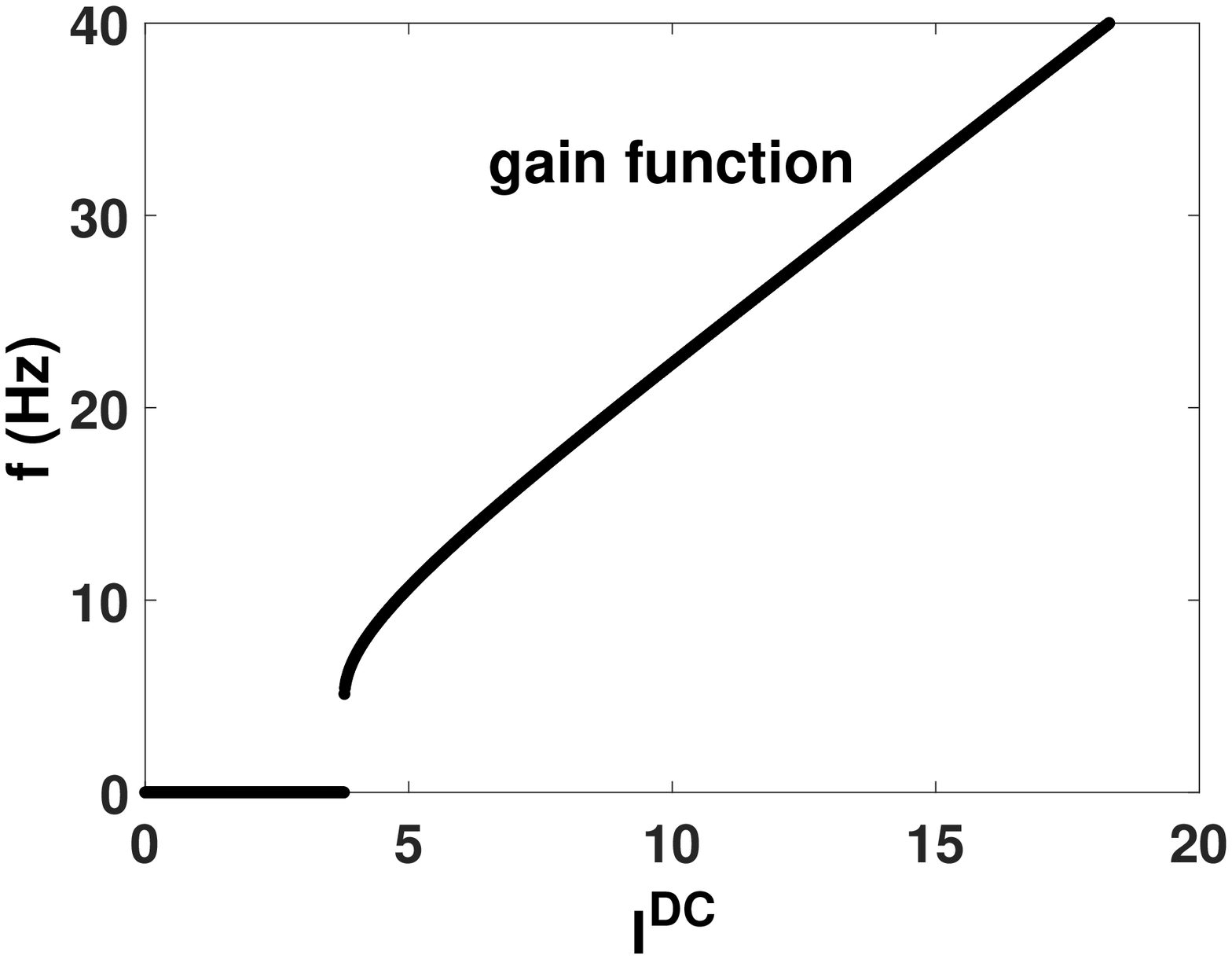}\label{fig1a}}
\subfigure[]{\includegraphics[width=0.48\textwidth,height=0.25\textwidth]{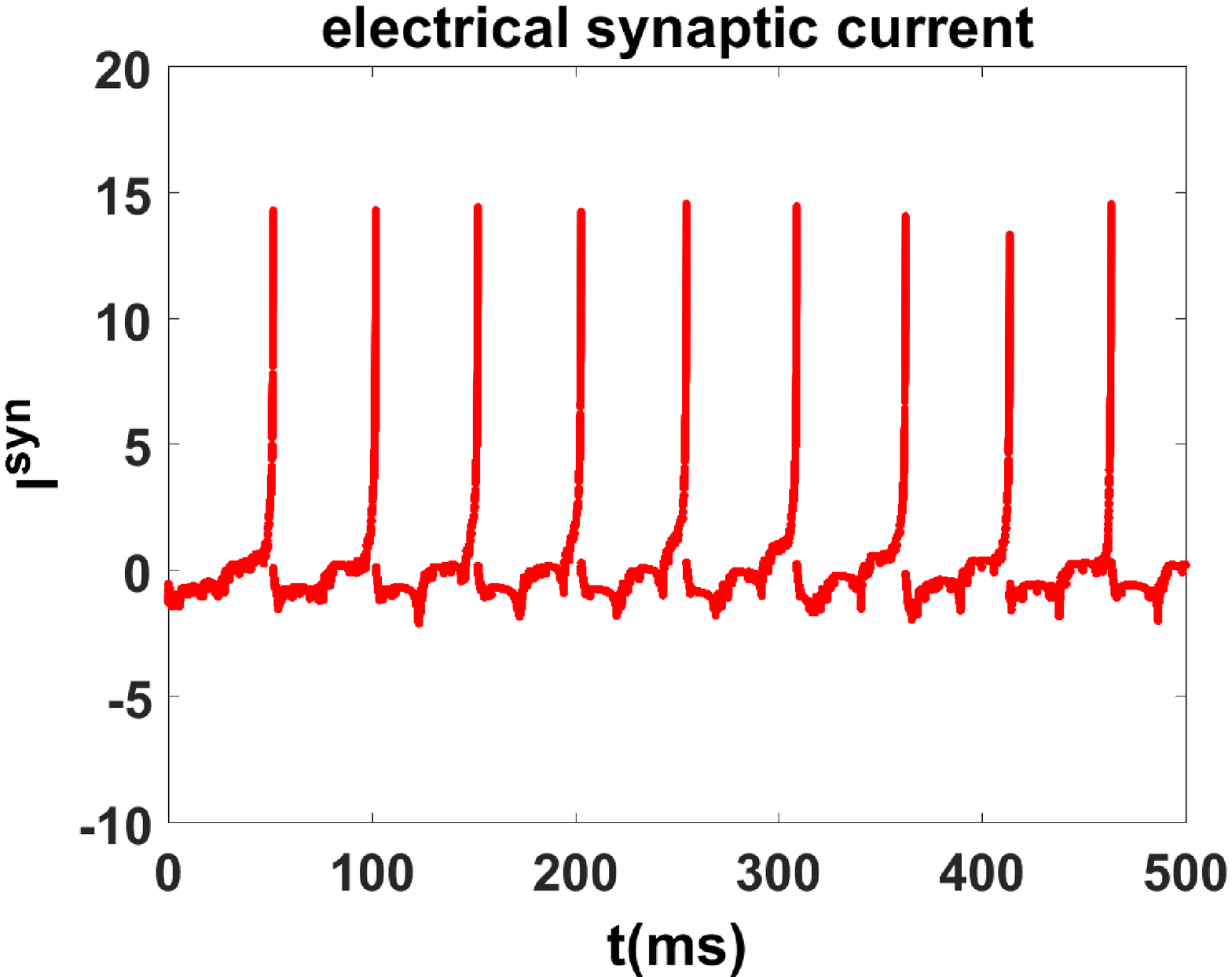}\label{fig1b}}
\subfigure[]{\includegraphics[width=0.48\textwidth,height=0.25\textwidth]{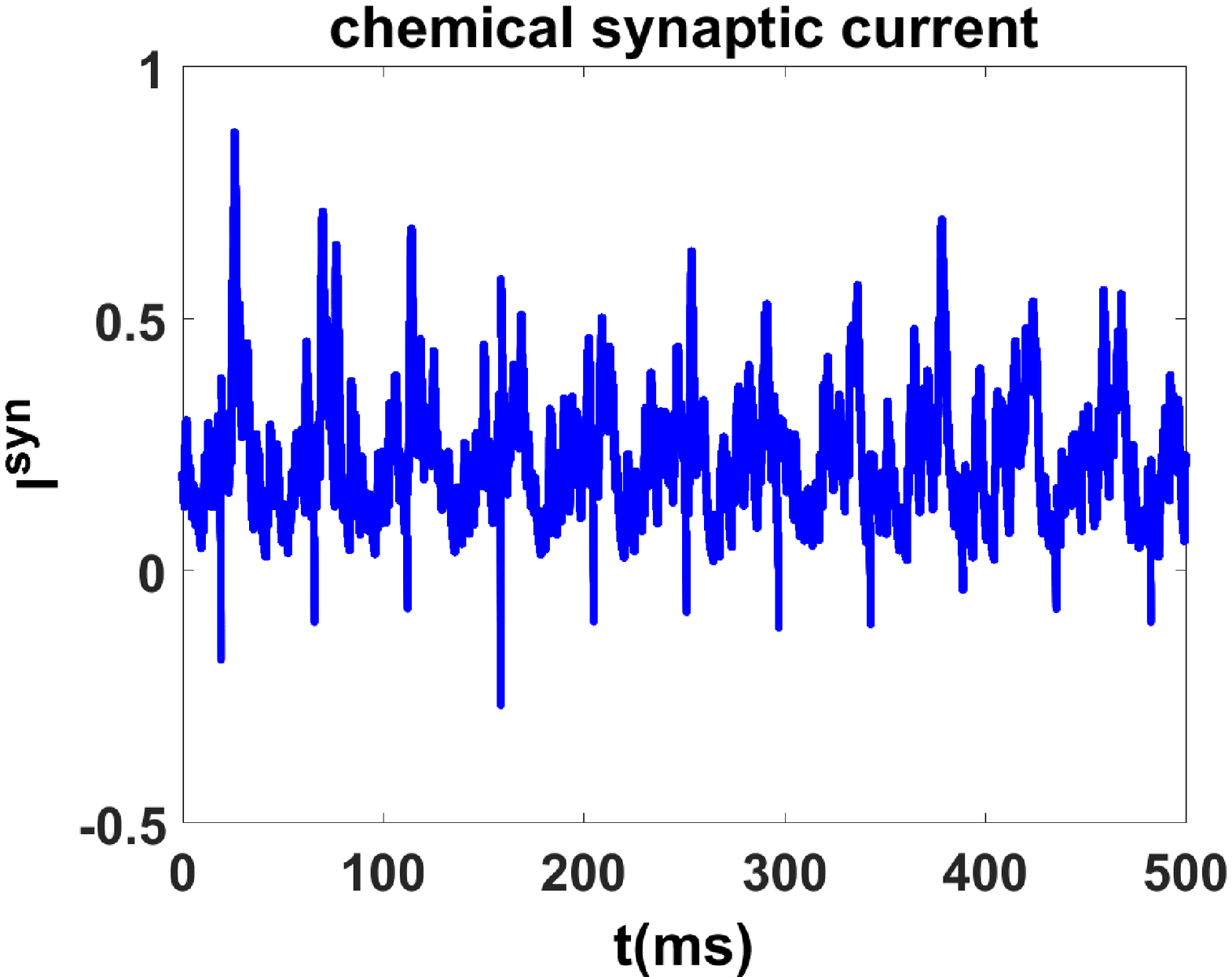}\label{fig1c}}
\end{center}
\caption{\small (a) Gain function of Izhikevich neuron which shows
the  dependence of firing rate of an uncoupled neuron to the
external current. (b) Electrical (and (c) Chemical) synaptic
current which an exemplary neuron in a network receives for
$g=0.15$. In this case the neurons of the network are
unsynchronized. $t=0$ indicates the beginning of stationary
state.} \label{fig1}
\end{figure}

In Fig.\ref{fig1a} we have plotted the gain function of regularly
spiking Izhikevich neuron. Gain function of a neuron shows the
dependence of firing rate on the external stimulating current
\cite{Gerstner}. It is seen that the neuron shows type II
excitability in this parameter regime and is capable of generating
regular spikes with a broad range of intrinsic frequencies from
theta-band to gamma-band. This range is more diverse than the
possible range of firing rates of HH neuron \cite{SGLee}.  Also
for an illustration, we have plotted the time dependence of the
electrical and chemical synaptic currents which an exemplary
neuron in a network receives from its neighbors at the beginning
of stationary state in Fig.\ref{fig1b} and \ref{fig1c},
respectively. Here $g=0.15$ and neurons of the circuit are
unsynchronized. We note that the pattern of electrical synaptic
current is very different from that of the chemical synaptic
current. For one thing, it is an order of magnitude stronger (15
vs 0.5). Secondly, they are dispersed and act more as a pulse as
opposed to fluctuating current due to chemical synapses. Thus it
is expected that electrical synapses have more impact on emergence
of synchronization in the system.
\begin{figure}[!htbp]
\begin{center}
\subfigure[]{\includegraphics[width=0.48\textwidth,height=0.25\textwidth]{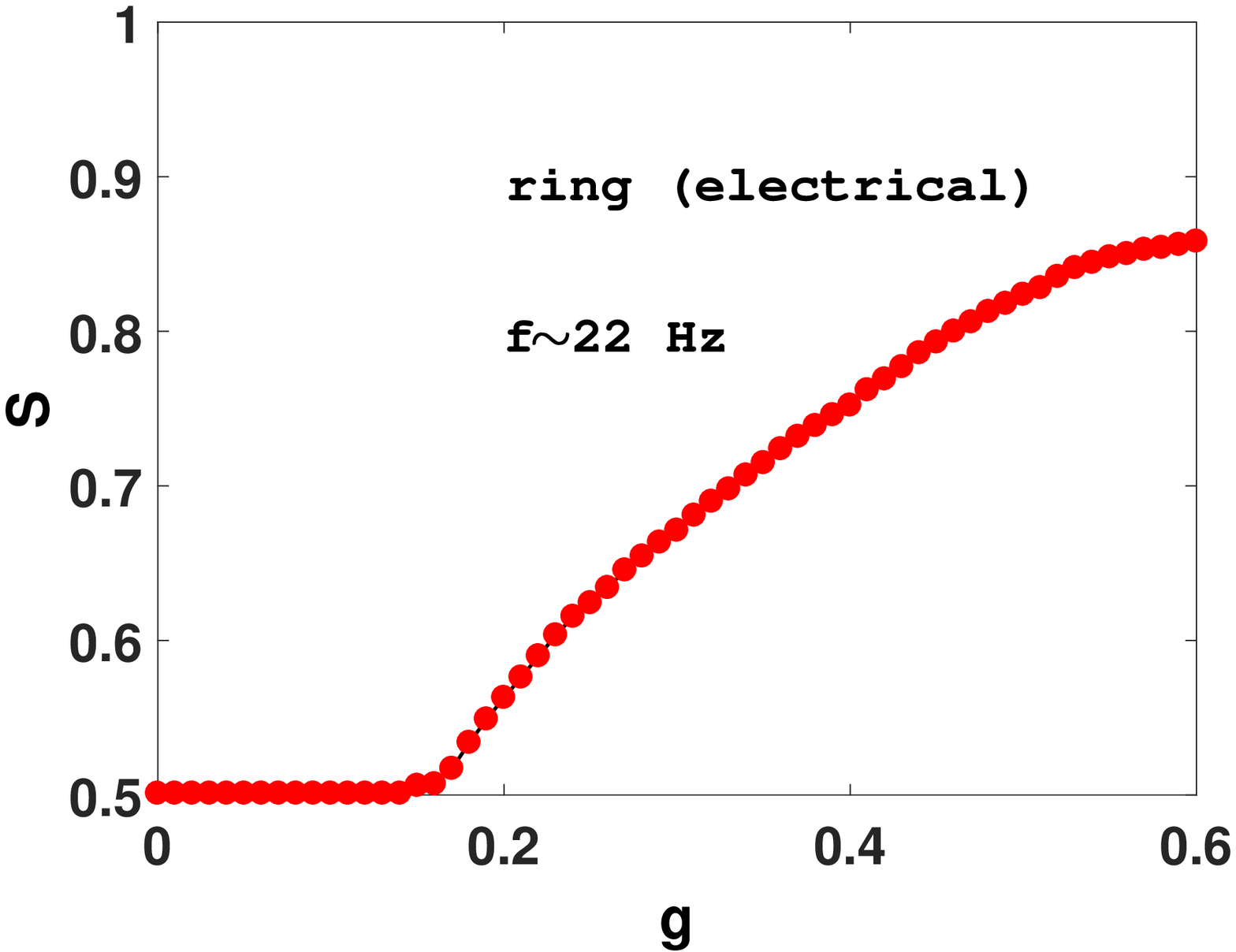}\label{fig2a}}
\subfigure[]{\includegraphics[width=0.48\textwidth,height=0.25\textwidth]{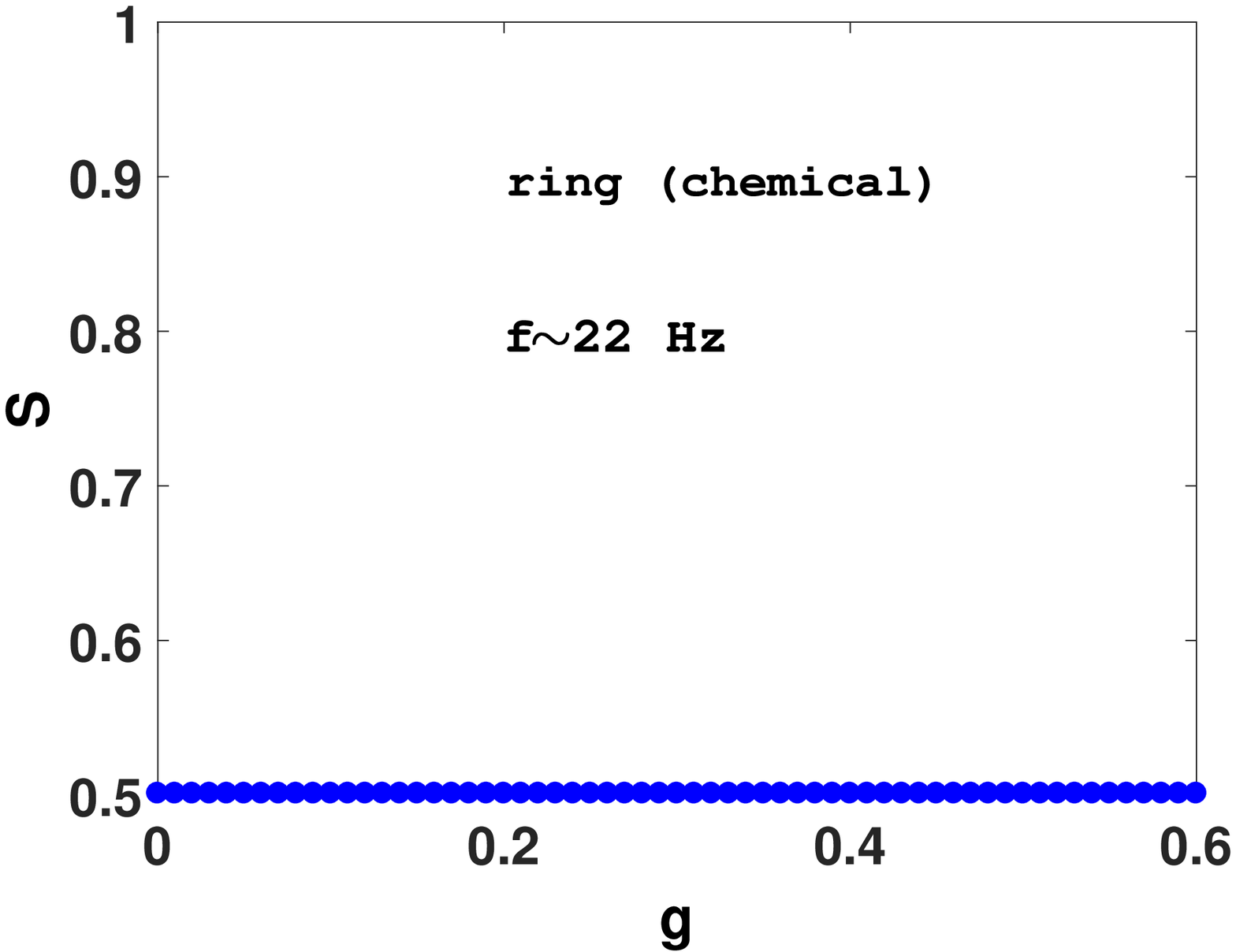}\label{fig2b}}
\subfigure[]{\includegraphics[width=0.48\textwidth,height=0.30\textwidth]{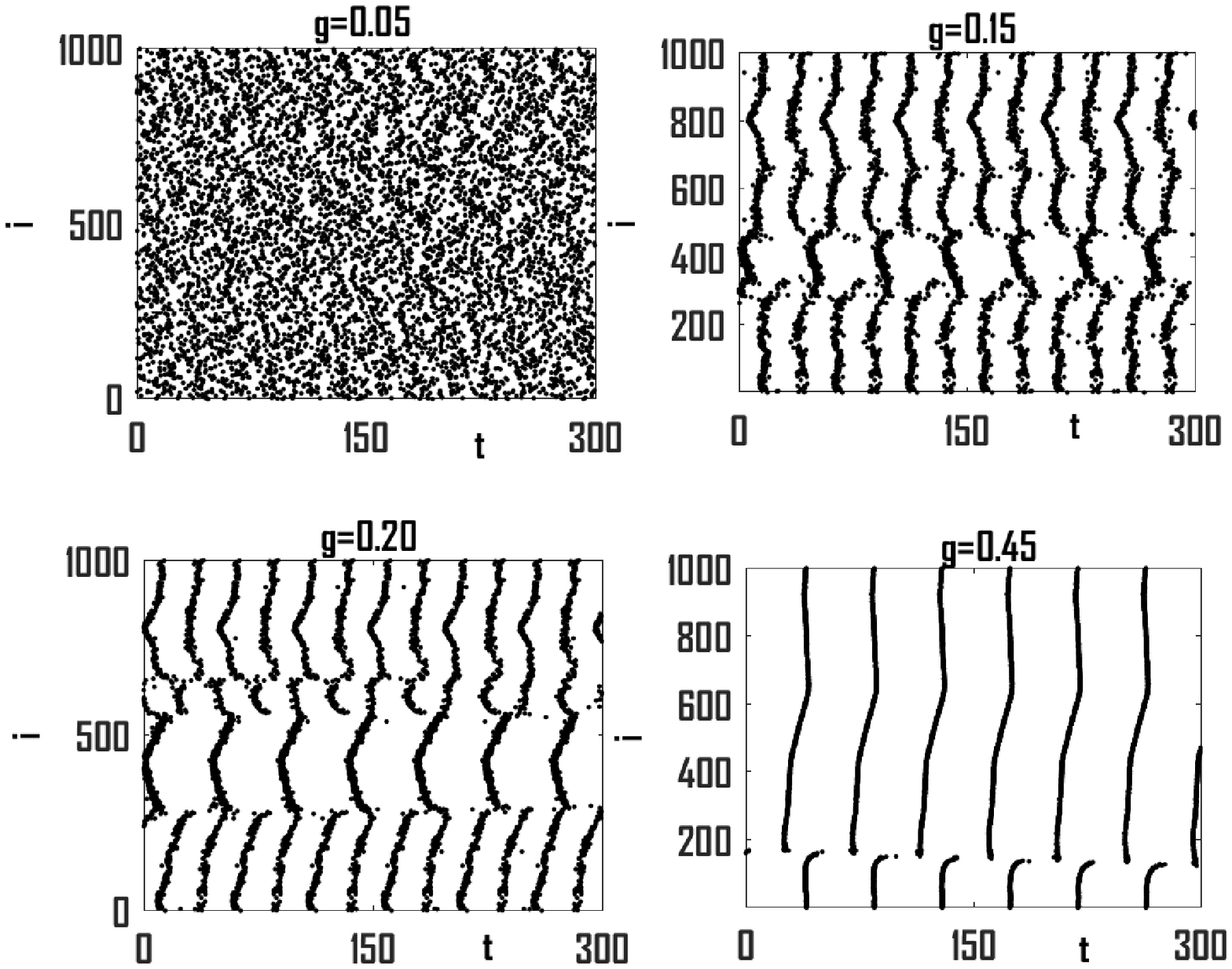}\label{fig2c}}
\subfigure[]{\includegraphics[width=0.48\textwidth,height=0.30\textwidth]{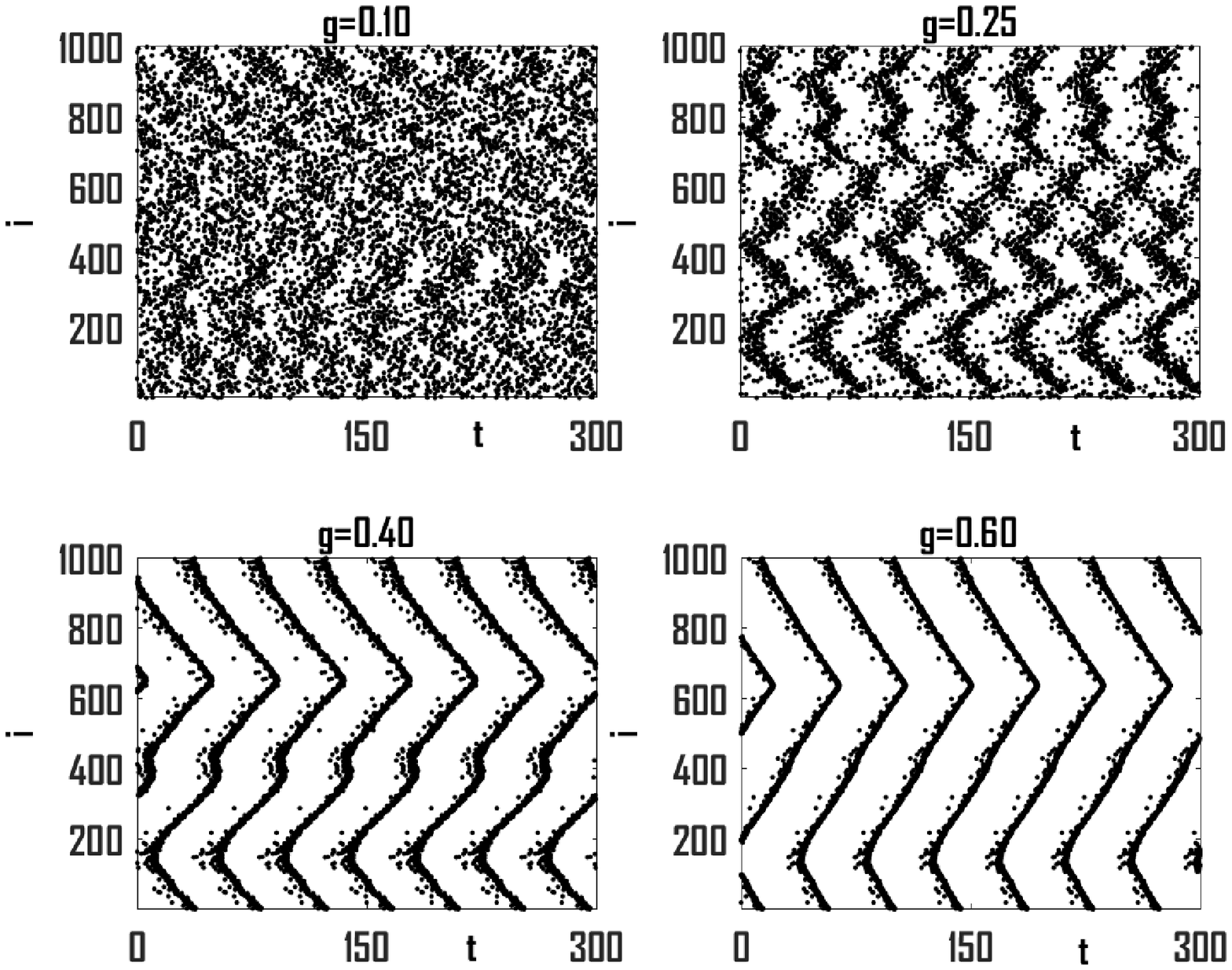}\label{fig2d}}
\end{center}
\caption{\small Synchronization diagram of Izhikevich neurons on a
regular ring for: (a) circuit with electrical synapses and (b)
circuit with chemical synapses. (c) Raster plots of the system of
panel (a) for four different values of $g$. (d) Raster plots of
the system in (b) for four different values of $g$. The mean
firing rate is $f{\simeq}22$ Hz here, and $t=0$ indicates the
beginning of stationary state. $t$ is measured in units of $ms$.} \label{fig2}
\end{figure}

Next we focus on synchronization transition in network models. We
construct networks of size $N=1000$ and coordination number
(average connectivity) $z=50$, unless otherwise stated. Also we
set the values of $I_i^{DC}$ so that the intrinsic firing rates
$f_i$ are in beta-band and have mean-value
$f={\langle}f_i{\rangle}{\simeq}22$ Hz, unless otherwise stated.
Synchronization diagrams for regular ring of Izhikevich neurons
with electrical and chemical synaptic interactions are illustrated
in Fig.\ref{fig2a} and \ref{fig2b}, respectively. It is observed
that the network with electrical synapses exhibits a continuous
transition to phase synchronization, while no transition occurs in
the network with chemical synapses. Investigation of raster plots
of the system with electrical synapses (Fig.\ref{fig2c}) reveals
that when synaptic interaction is weak, neurons spike
out-of-phase. Note that the mean firing rate is $f{\simeq}22$ Hz,
and therefore each neuron should fire about seven times in the 300
ms window that is illustrated here. Increasing $g$ slightly, leads
to two neural groups each of which contains neurons that spike
partially coherently but the members of two groups spike
anti-phase with respect to each other. See $g=0.15$ in
Fig.\ref{fig2c}. When we increase $g$, the phases of a number of
members in one group gradually match the phases of the members of
the other group. Hence the order parameter of synchronization
increases continuously from $S=0.5$ to higher values and the
neural network exhibits a continuous transition to phase
synchronization. In case of ring with chemical synapses, no
anti-phase groups form. Since neuronal interactions are local,
increasing synaptic strength, leads to the formation of wave-like
pattern in order of neuronal spikes (Fig.\ref{fig2d}). Although
increasing $g$ enhances local coherence in neuronal oscillations
and each neuron has a small phase lag with adjacent neurons in the
circuit, there exists no global order in the network and
$S{\simeq}0.5$, see Fig.\ref{fig2d}. We cannot increase $g$ to
arbitrary large values, as after a certain value (near
$g{\simeq}0.6$) neurons start bursting instead of regular spiking
\cite{Kim}. We therefore conclude that spiking Izhikevich neurons
with chemical synapses with purely local interactions lead to
local order without any long-range order necessary for a phase
transition. However, the effect of long-range interaction may
change this picture.

\begin{figure}[!htbp]
\begin{center}
\subfigure[]{\includegraphics[width=0.53\textwidth,height=0.25\textwidth]{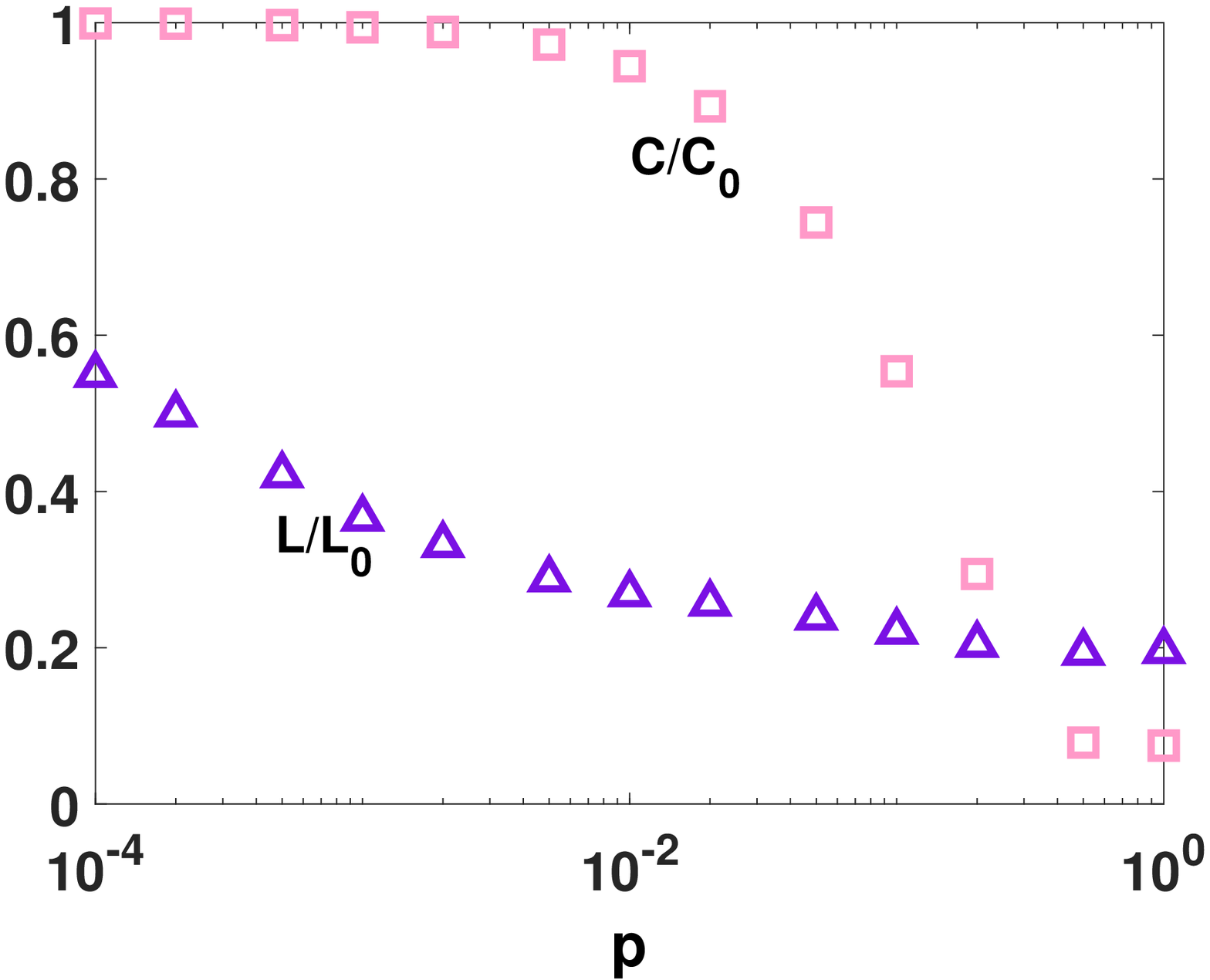}\label{fig3a}}
\subfigure[]{\includegraphics[width=0.48\textwidth,height=0.25\textwidth]{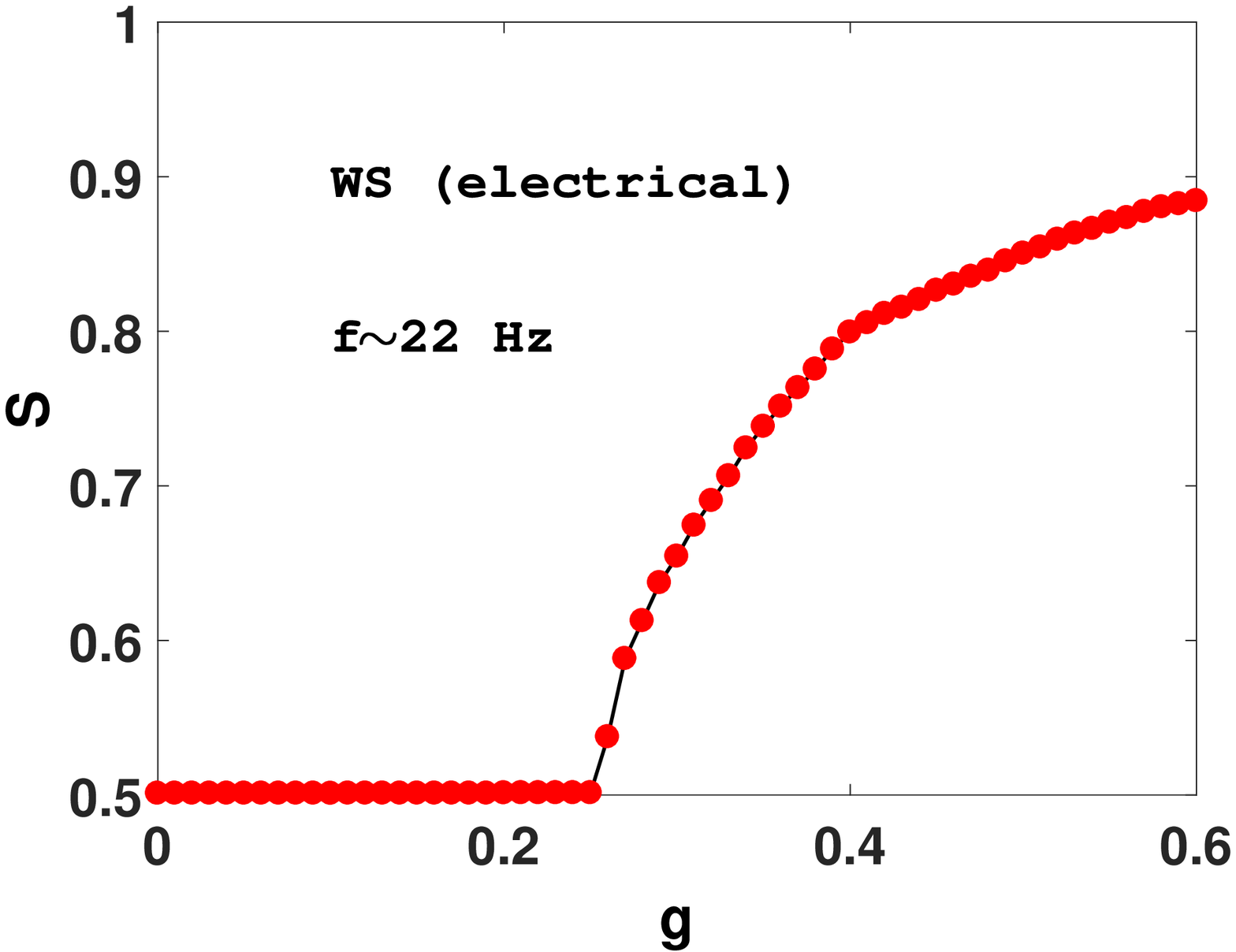}\label{fig3b}}
\subfigure[]{\includegraphics[width=0.48\textwidth,height=0.25\textwidth]{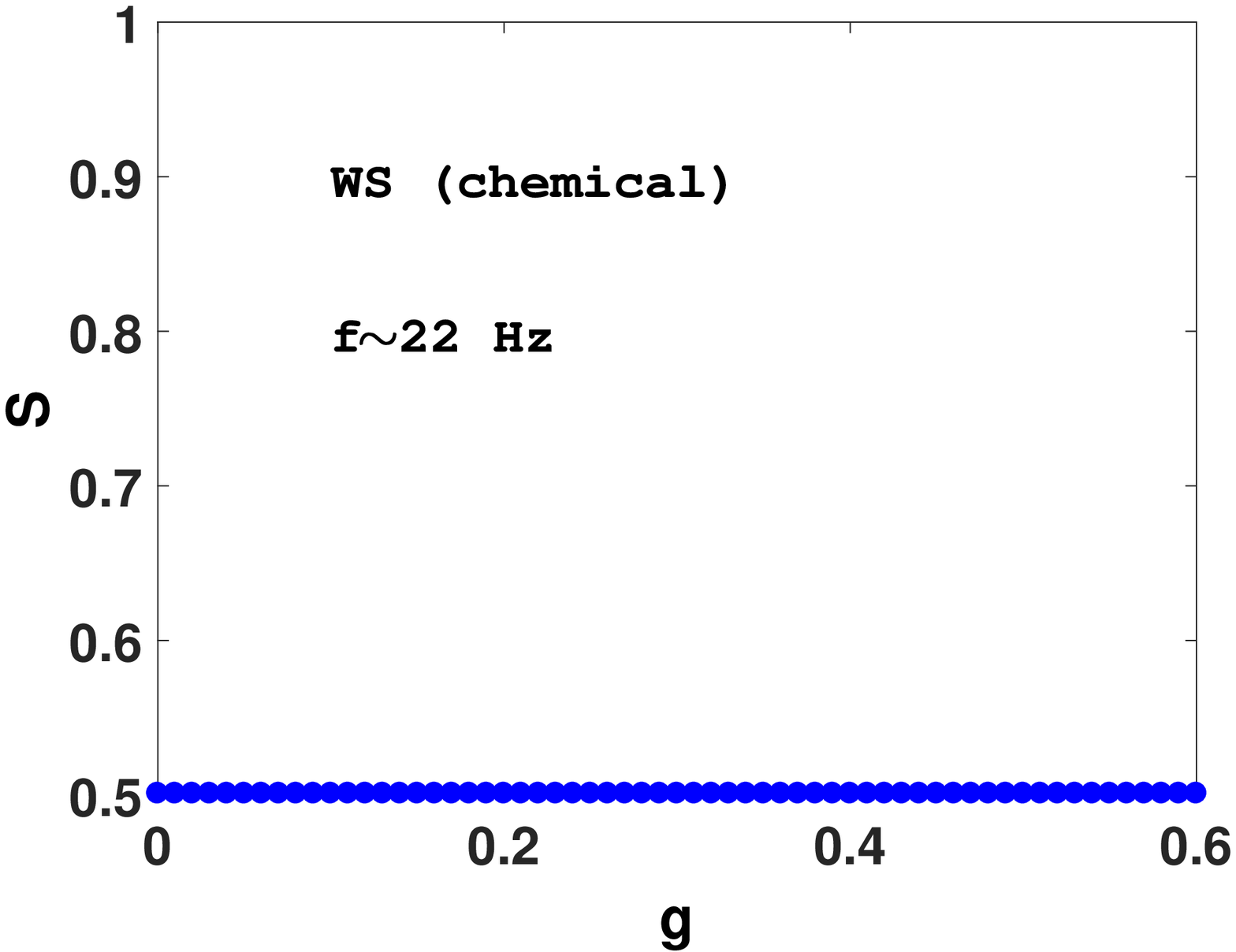}\label{fig3c}}
\end{center}
\caption{\small (a) Dependence of clustering coefficient $C$ and
average path length $L$ on rewiring probability $p$ in WS networks
with $N=1000$ and $z=50$. $C$ and $L$ are normalized with $C_0$
and $L_0$ which are the clustering coefficient and average path
length of a regular ring ($p=0)$, respectively. (b) and (c)
Synchronization diagram of Izhikevich neurons on WS networks with
$p=0.01$, for electrical and chemical synaptic interactions. The
mean firing rate is $f{\simeq}22$ Hz here.} \label{fig3}
\end{figure}

To examine this we consider transition to phase synchronization in
Watts-Strogatz (WS) small-world networks \cite{Watts}. In
Fig.\ref{fig3a} we have plotted the variation of clustering
coefficient $C$ and average distance $L$ when we rewire the
previous ring with different probabilities $p$ and found that for
$p=0.01$ the resulting network has significant small-world effect
and clustering coefficient, simultaneously. Figs.\ref{fig3b} and
\ref{fig3c} show the synchronization diagrams of Izhikevich
neurons with electrical and chemical synapses in WS networks with
$p=0.01$. It is seen that the resultant synchronization diagrams
are similar to $S-g$ plots of regular ring except for a different
(larger) transition point in circuits with electrical synapses.
Investigation of the raster plots of WS neural networks (not
shown) reveals that the underlying reason for the observed
synchronization transitions is exactly the same as the reason
explained for regular ring above. Since our regular ring and WS
network have approximately the same value of $C$ but distinctly
different values of $L$, similarity of the transitions which they
produce indicates that the clustering coefficient (and not the
small-world effect) is the main topological factor that plays an
important role in the resulting transitions.
\begin{figure}[!htbp]
\begin{center}
\subfigure[]{\includegraphics[width=0.48\textwidth,height=0.25\textwidth]{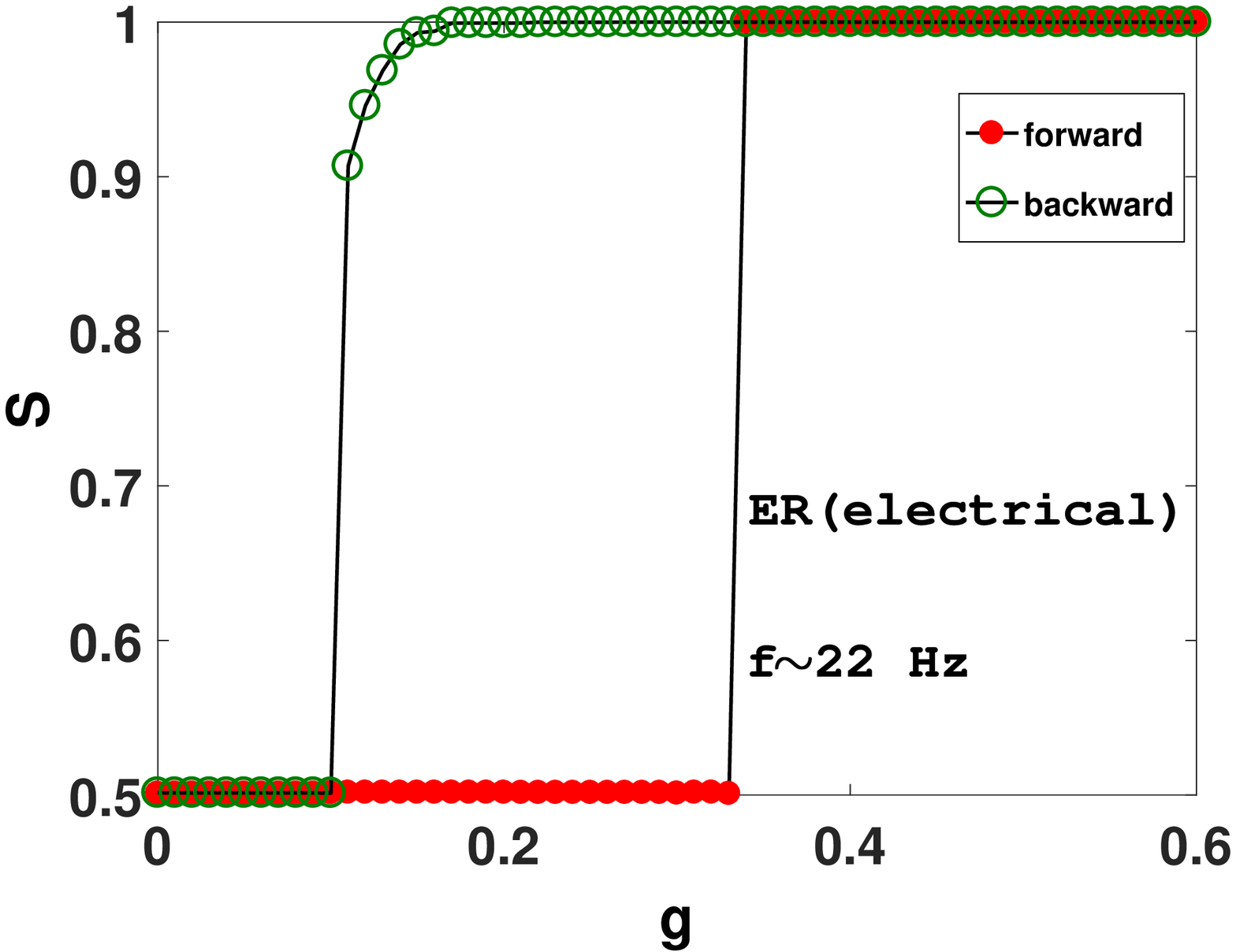}\label{fig4a}}
\subfigure[]{\includegraphics[width=0.48\textwidth,height=0.25\textwidth]{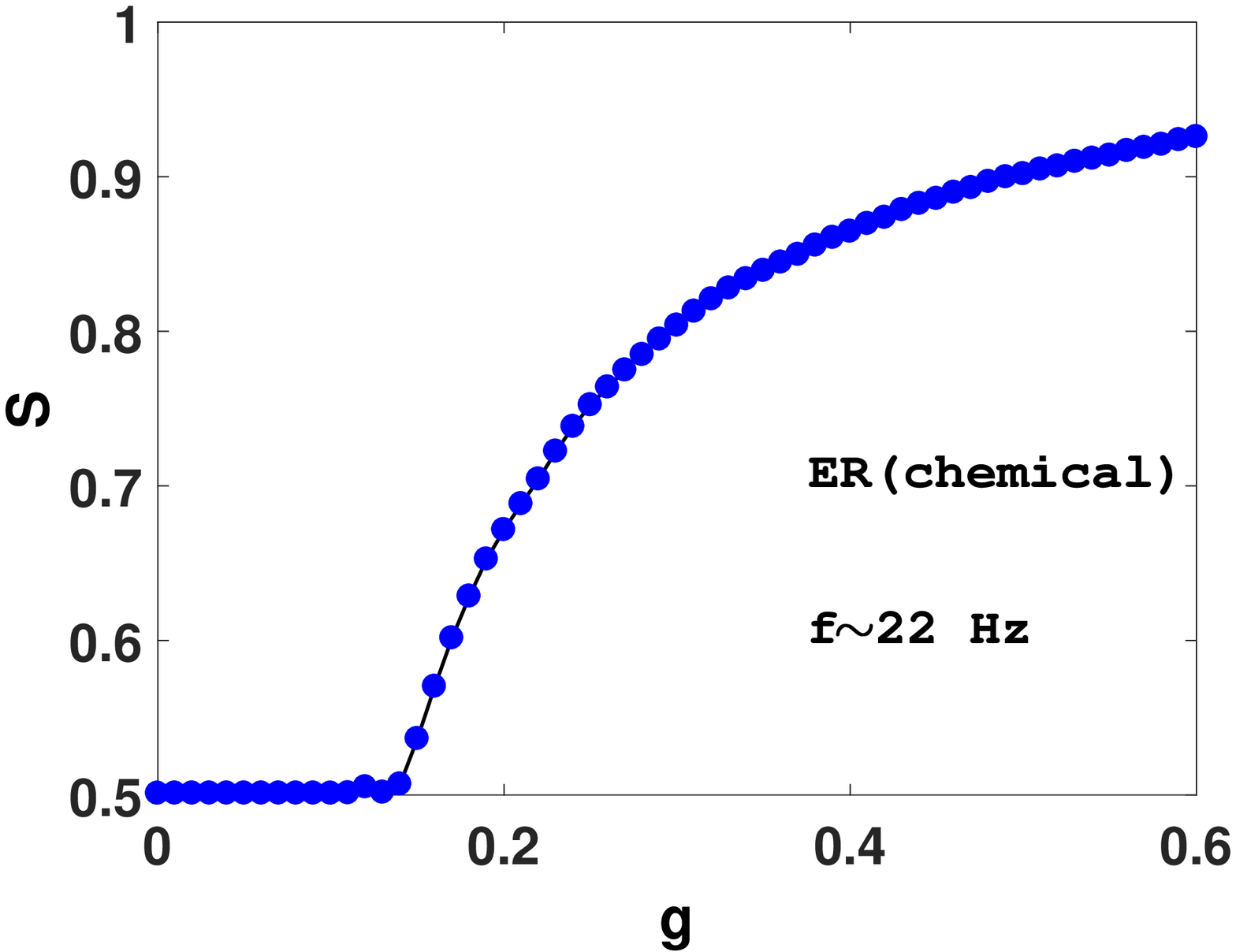}\label{fig4b}}
\subfigure[]{\includegraphics[width=0.48\textwidth,height=0.30\textwidth]{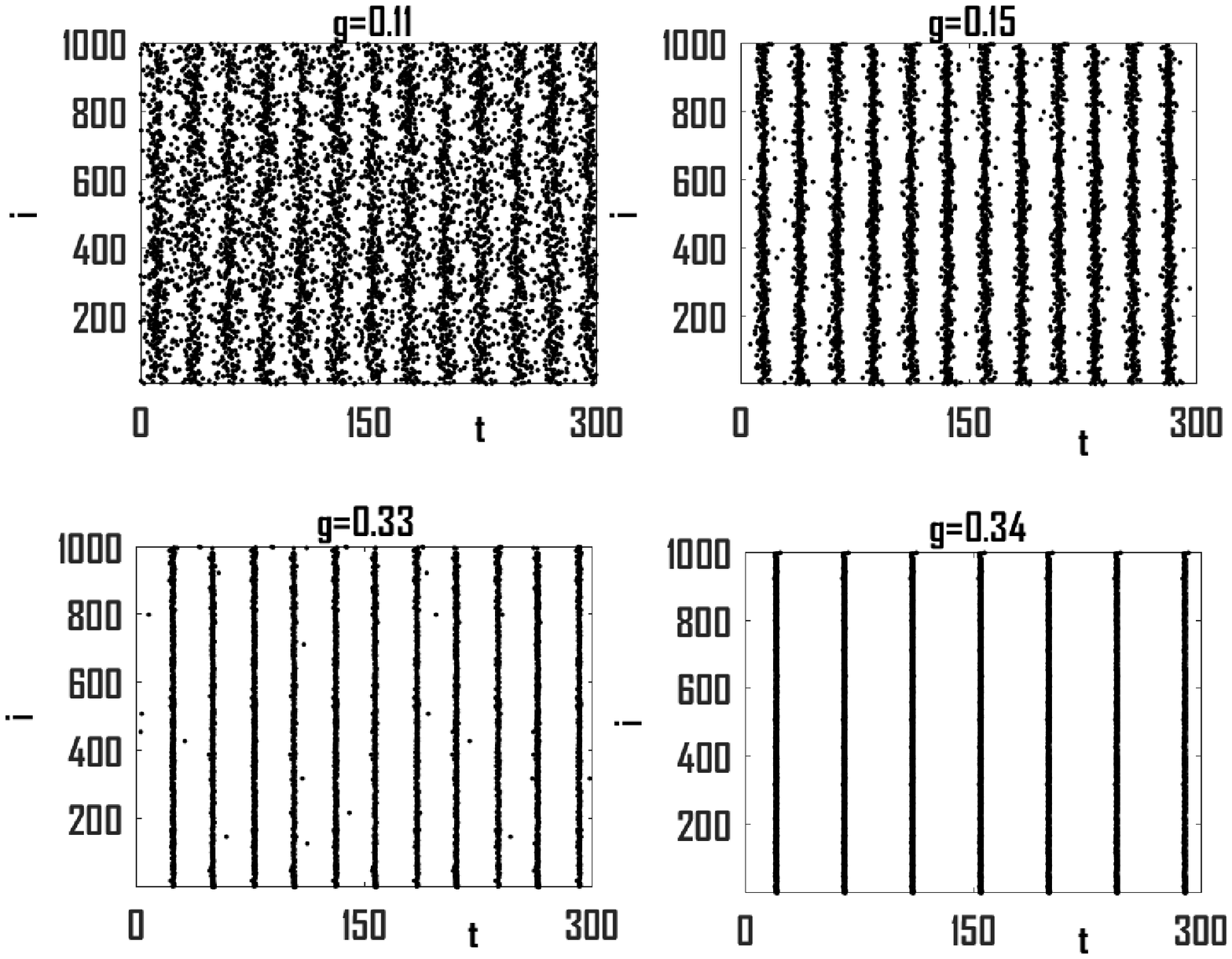}\label{fig4c}}
\subfigure[]{\includegraphics[width=0.48\textwidth,height=0.30\textwidth]{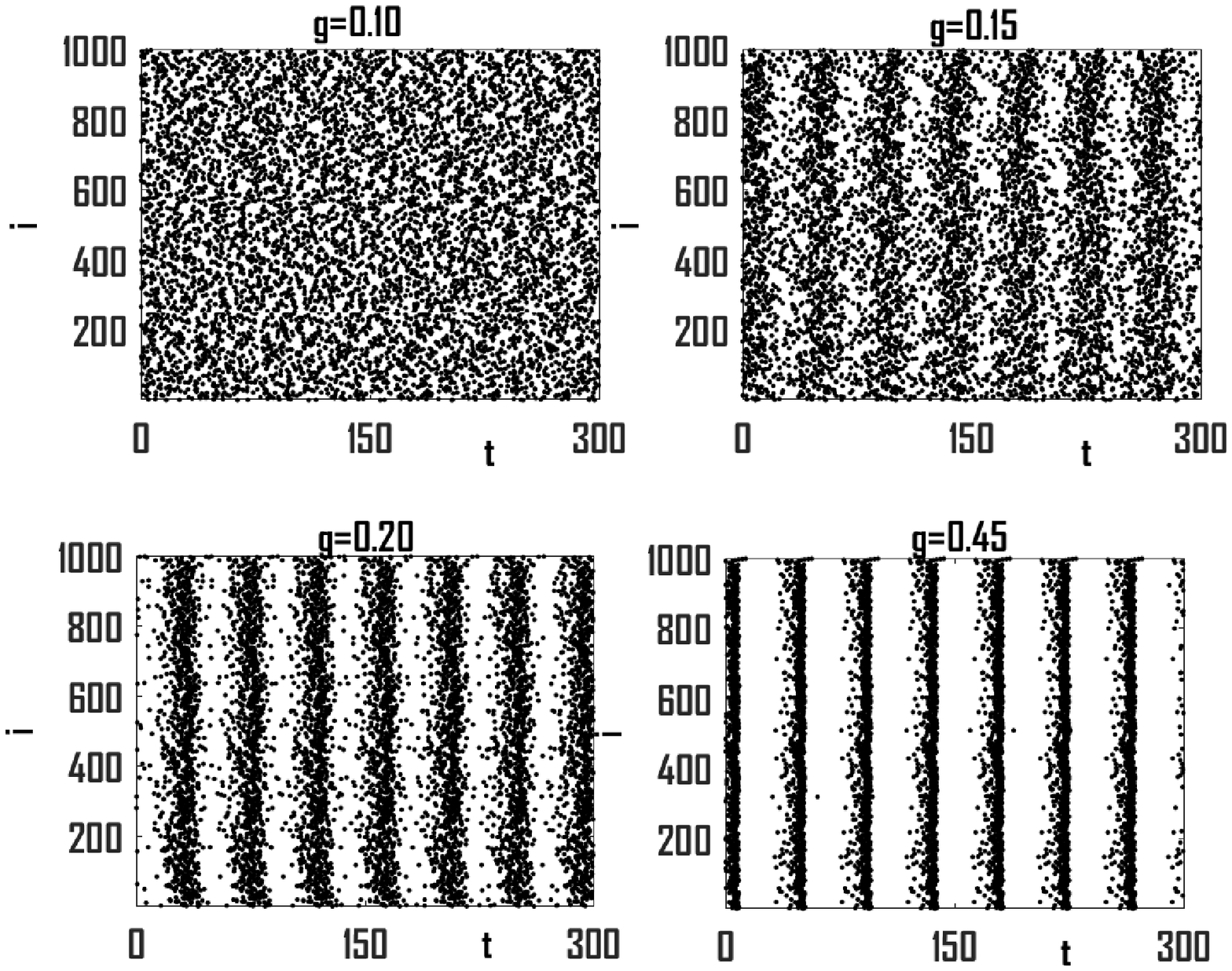}\label{fig4d}}
\end{center}
\caption{\small  Synchronization diagram of Izhikevich neurons on
ER network: (a) $S-g$ plot in forward and backward evolution of
the system with electrical synapses. (b) $S-g$ plot for the system
with chemical synapses. (c) Raster plots for the system with
electrical synapses in forward direction. (d) Raster plots of the
system with chemical synapses. The mean firing rate is
$f{\simeq}22$ Hz and $t=0$ indicates the beginning of stationary
state. $t$ is measured in units of $ms$.} \label{fig4}
\end{figure}

In order to examine the role of clustering coefficient further, we
investigate more random topologies, viz, Erdos-Renyi (ER) network
\cite{Erdos} and scale-free (SF) network with small average path
length and negligible clustering coefficient \cite{Gros}. Both
these networks have random structures with ER being homogeneous
and SF being heterogenous exhibiting hubs which are thought to be
to play an important role in synchronization phenomena
\cite{Gardenes2011}. Also, both these networks exhibit small-world
effect, while the existence of hubs in SF networks leads to a
relatively smaller $L$ for a fixed $z$ and $N$. Dependence of
order parameter $S$ on coupling strength $g$ for an ER network of
Izhikevich neurons with electrical synapses is illustrated in
Fig.\ref{fig4a}. The network exhibits a first-order or explosive
transition to phase synchronization, with a large hysteresis loop,
as neurons spike with beta rhythms. Note that the transition is
truly explosive as $S$ jumps directly to its maximum value
immediately at the transition point, indication full synchrony in
the network. Explosive synchronization is a novel phenomenon and
has attracted much attention recently. Different mechanisms have
been reported for generation of such type of synchronization
transition so far
\cite{Gardenes2011,Leyva,Zhang2015,Ji2013,Skardal2014}, and the
key role played by heterogeneity has been in focus in this regard.

Seeking the underlying reason of this explosive transition, we
investigate raster plots of this neural circuit for different
values of $g$. Four such raster plots  for forward evolution of
the system are shown in Fig.\ref{fig4c}. We find that neurons
spike out-of-order initially. As $g$ is increased slightly, the
neurons in the system are organized into two distinct groups in
which members of each group spike almost coherently, as they
oscillate anti-phase with the other group. Further increase of $g$
regulates neuronal phases in each group but the phase lag between
two groups remain robust. Therefore there exists no global phase
coherence in the system and $S=0.5$, see $g=0.33$ in
Fig.\ref{fig4c}.  There exists a transition point for which these
two anti-phase groups abruptly join together leading to complete
phase coherence. Hence the order parameter suddenly jumps from
$S=0.5$ to $S=1$, see Fig.\ref{fig4a} and $g=0.34$ in
Fig.\ref{fig4c}.

When interaction among neurons is mediated via the softer chemical
synapses (Fig.\ref{fig4b}), anti-phases groups do not form in the
neural network. Since the clustering coefficient of  ER network is
negligible ($C=0.054$ here) and long-range interaction is
significant, wave-like patterns in neuron spikes do not appear.
See raster plots in Fig.\ref{fig4d}. The gradual increase of $g$
subsequently leads to gradual increase in global order in the
system leading to a continuous transition at which global order
appear in the system, see Fig.\ref{fig4d}. Further increase of $g$
leads to increase of $S$ as more and more neurons align their
phases. Therefore Izhikevich neurons with mean firing rate
$f{\simeq}22$ Hz produce continuous transition to phase
synchronization when they interact via chemical synapses on an ER
network.

\begin{figure}[!htbp]
\begin{center}
\subfigure[]{\includegraphics[width=0.48\textwidth,height=0.25\textwidth]{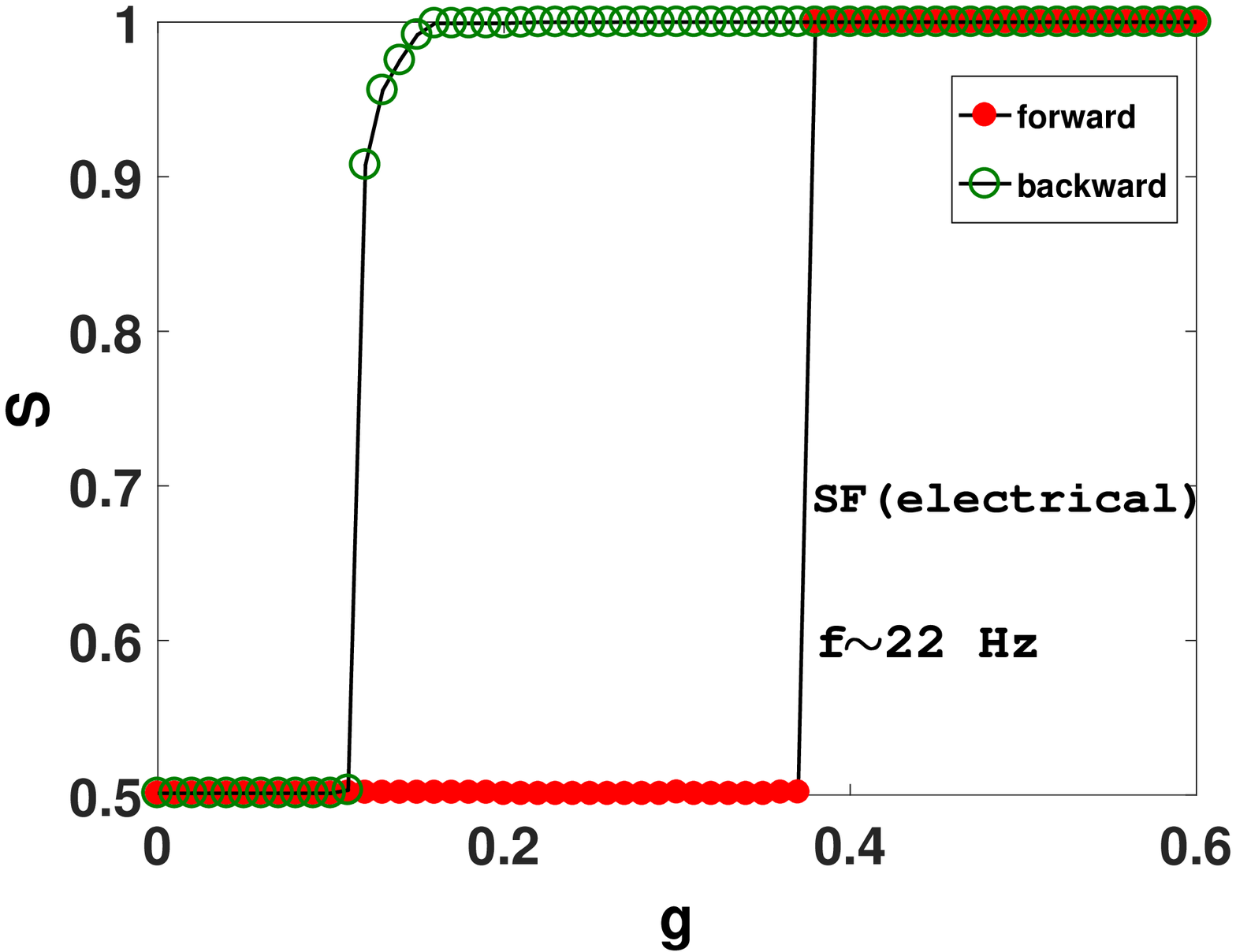}\label{fig5a}}
\subfigure[]{\includegraphics[width=0.48\textwidth,height=0.25\textwidth]{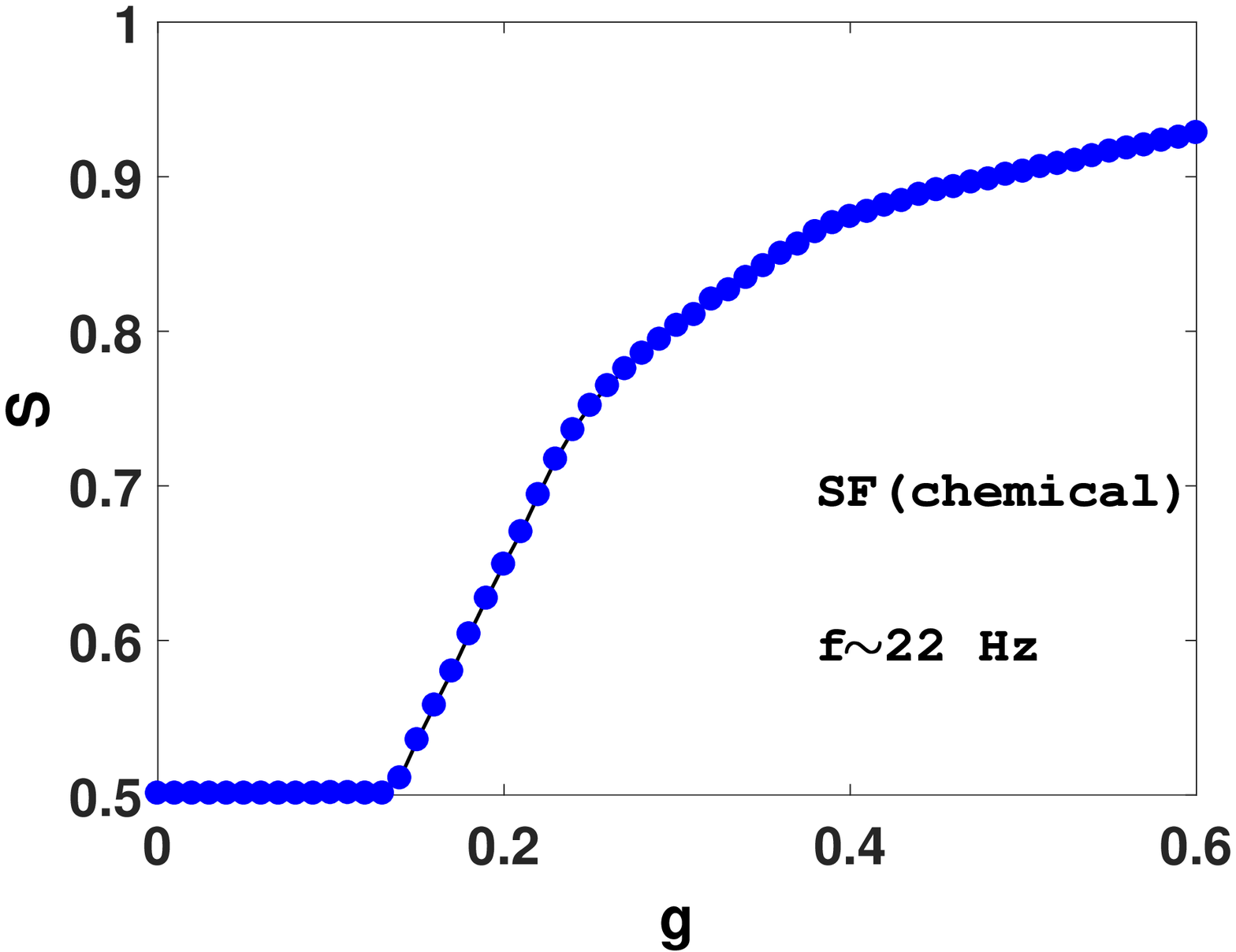}\label{fig5b}}
\end{center}
\caption{\small Synchronization diagram of Izhikevich neurons on
SF networks with $N=1000$, $z=20$ and degree distribution function
$P(k){\sim}k^{-\gamma}$ with $\gamma=3$. (a) Phase transition
diagram for the system with electrical synapses in forward and
backward evolution of the system. (b) Phase transition diagram for
the system with chemical synapses. The mean firing rate is
$f{\simeq}22$ Hz.} \label{fig5}
\end{figure}

Next, we ask whether heterogeneity in SF networks can change the
picture obtained from ER networks above.  Fig.\ref{fig5} displays
synchronization diagrams of Izhikevich neurons on SF networks.
Here we have generated uncorrelated SF networks \cite{Catanzaro}
with coordination number $z=20$ and degree distribution function
$P(k){\sim}k^{-\gamma}$ with $\gamma=3$. Smaller $z$ is necessary
here in order to give real meaning to heterogeneity needed in our
study for SF networks.  Note, that despite using smaller $z$ the
network still displays significant small-world effect and small
clustering coefficient, see Table I. $S-g$ plots of Izhikevich
neurons with mean firing rate $f{\simeq}22$ Hz on SF networks with
electrical and chemical synapses are illustrated in
Fig.\ref{fig5a} and Fig.\ref{fig5b}, respectively. Interestingly,
it is observed that the resulting synchronization diagrams are
essentially exactly the same as the results obtained for ER
network. We therefore conclude that the existence of hubs does not
have a significant effect in the synchronization pattern in the
parameter regime we have focused for Izhikevich neurons.
Furthermore, the dramatic change in clustering coefficient of
random networks (ER or SF) lead to decidedly different
synchronization pattern when compared to clustered networks such
as regular ring or WS network.

\begin{figure}[!htbp]
\begin{center}
\subfigure[]{\includegraphics[width=0.48\textwidth,height=0.25\textwidth]{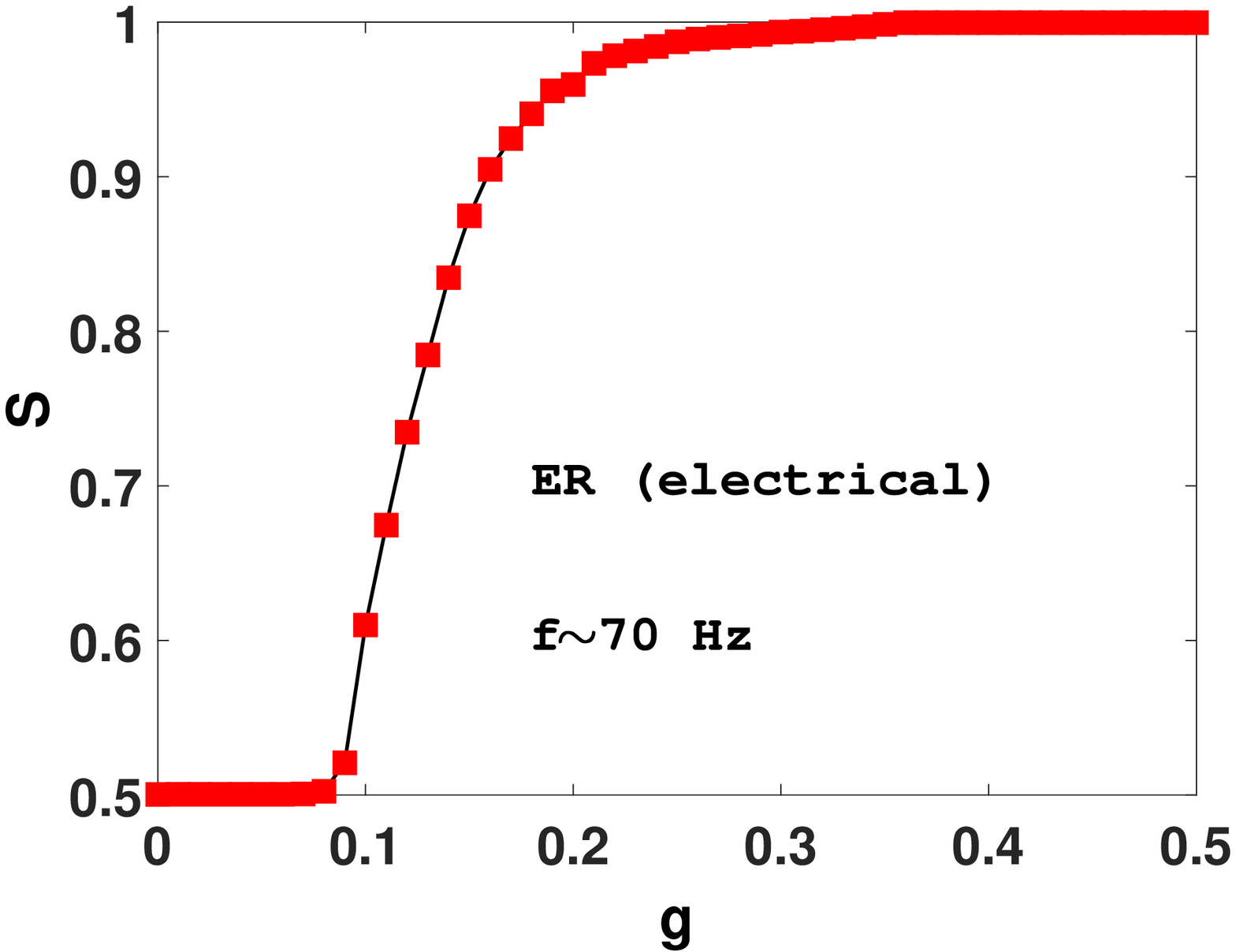}\label{fig6a}}
\subfigure[]{\includegraphics[width=0.48\textwidth,height=0.25\textwidth]{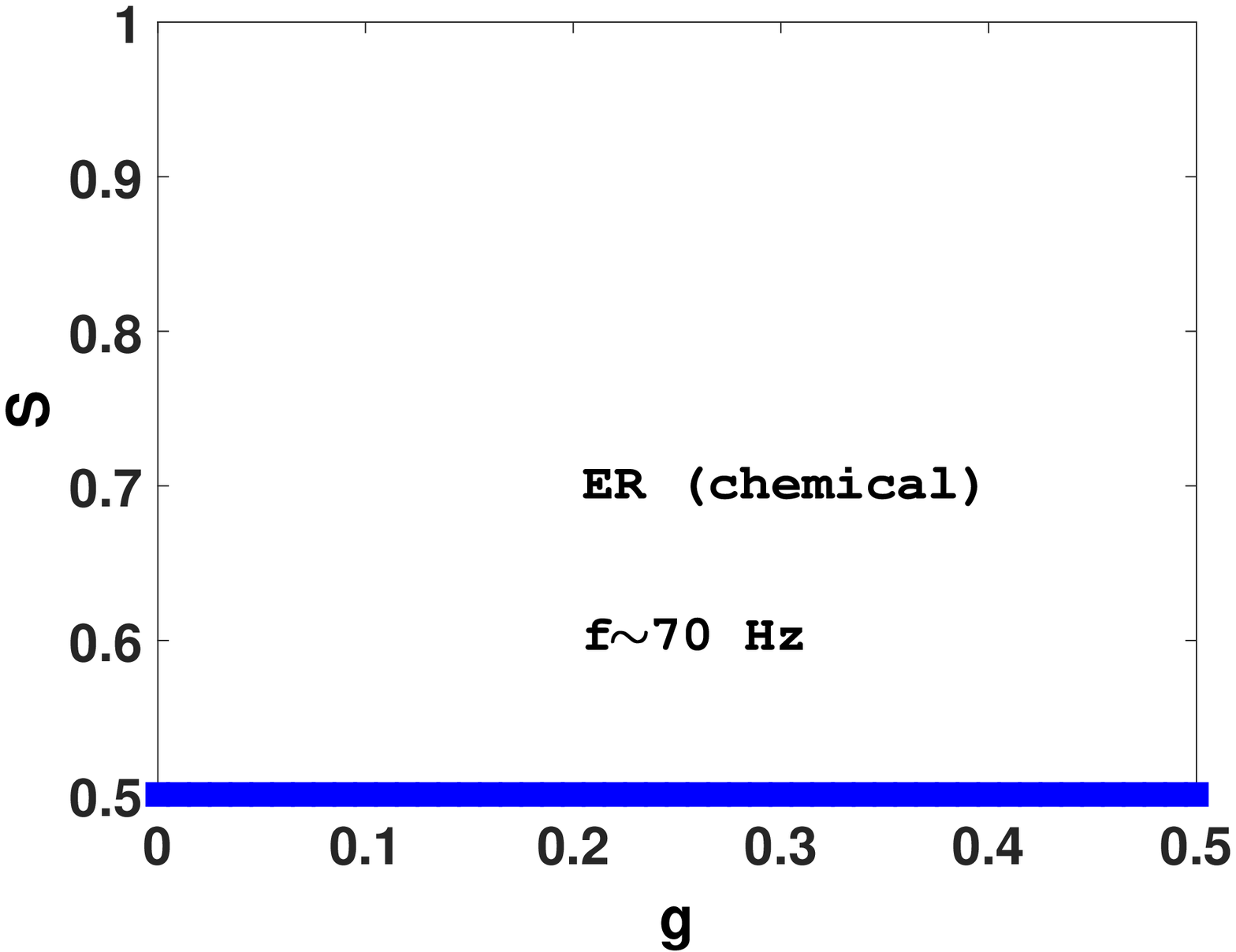}\label{fig6b}}
\subfigure[]{\includegraphics[width=0.54\textwidth,height=0.35\textwidth]{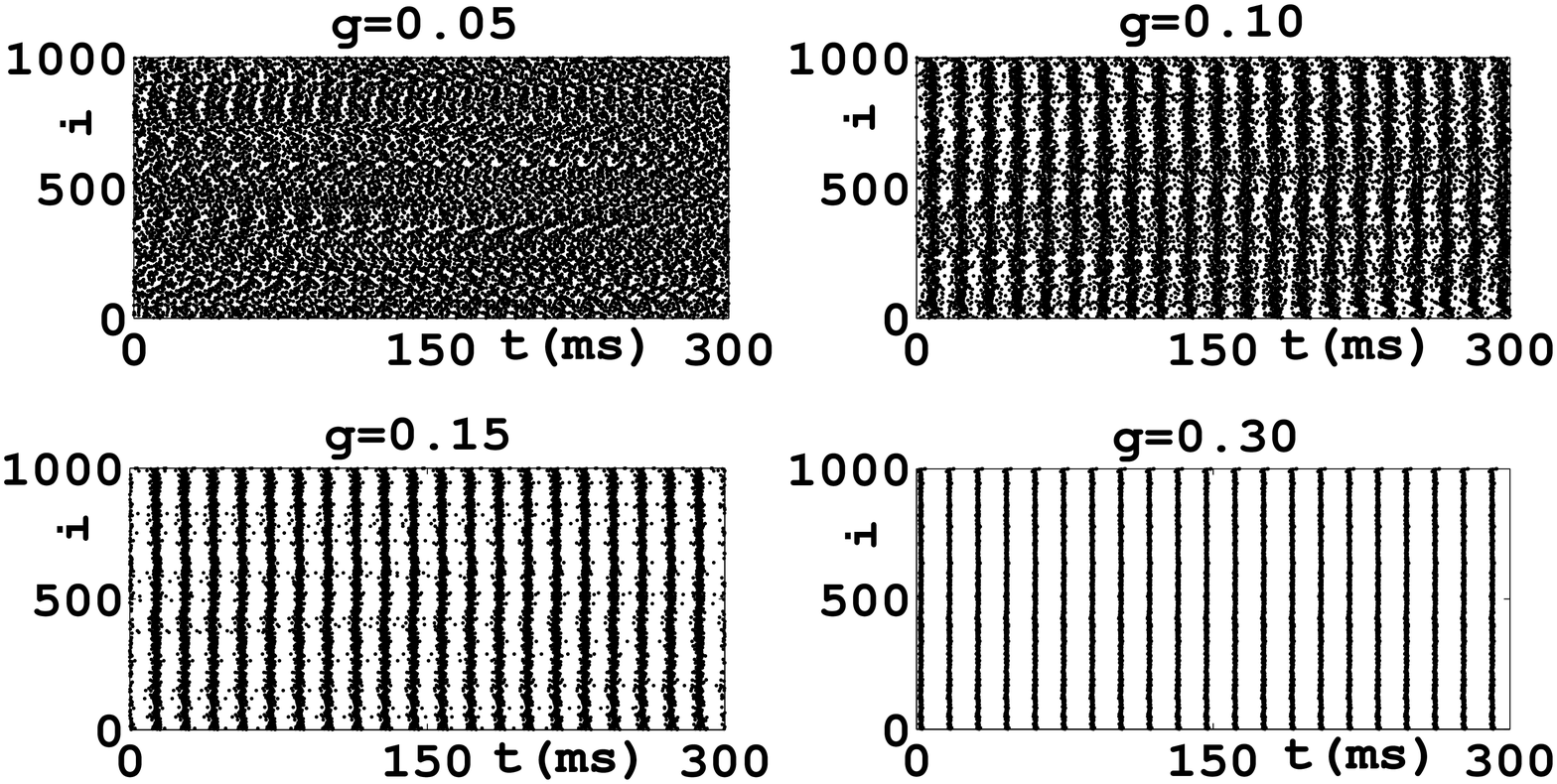}\label{fig6c}}
\end{center}
\caption{\small Synchronization diagram of Izhikevich neurons with
spiking frequencies in the high gamma band on ER networks: Phase
transition diagram for the system with (a) electrical synapses and
(b) chemical synapses. (c) Raster plots for the system with
electrical synapses. The mean firing rate is $f{\simeq}70$ Hz and
$t=0$ indicates the beginning of stationary state.} \label{fig6}
\end{figure}

When synchronization transition is studied in a population of
phase oscillators such as Kuramoto model, results are independent
of the mean value of frequency distribution. Therefore we can
switch to a rotating frame of reference where the mean value of
frequencies is zero \cite{Gros}. In contrast we found that the
resulting synchronization transitions which we obtain in neural
circuits depend on the mean frequency of firing. For example, in
Fig.\ref{fig6} we have illustrated synchronization diagrams and
raster plots of Izhikevich neurons with firing frequencies in high
gamma band ($f={\langle}f_i{\rangle}{\simeq}70$ Hz) on ER network.
Comparing these results with plots of Fig.\ref{fig4}, it is
observed that while neurons oscillate with high gamma rhythms,
electrical synapses lead to a continuous transition to phase
synchronization (rather than the explosive transition in beta
frequencies) and chemical synapses do not lead to any
synchronization in the system (as opposed to a continuous
transition in beta frequencies). Also investigation of raster
plots shows that while neurons fire with high gamma frequencies,
interactions via electrical synapses do not results in anti-phase
synchronization. (Compare Fig.\ref{fig4c} and Fig.\ref{fig6c}).
This is a curious result that needs further investigation.

\begin{figure}[!htbp]
\begin{center}
\subfigure[]{\includegraphics[width=0.48\textwidth,height=0.25\textwidth]{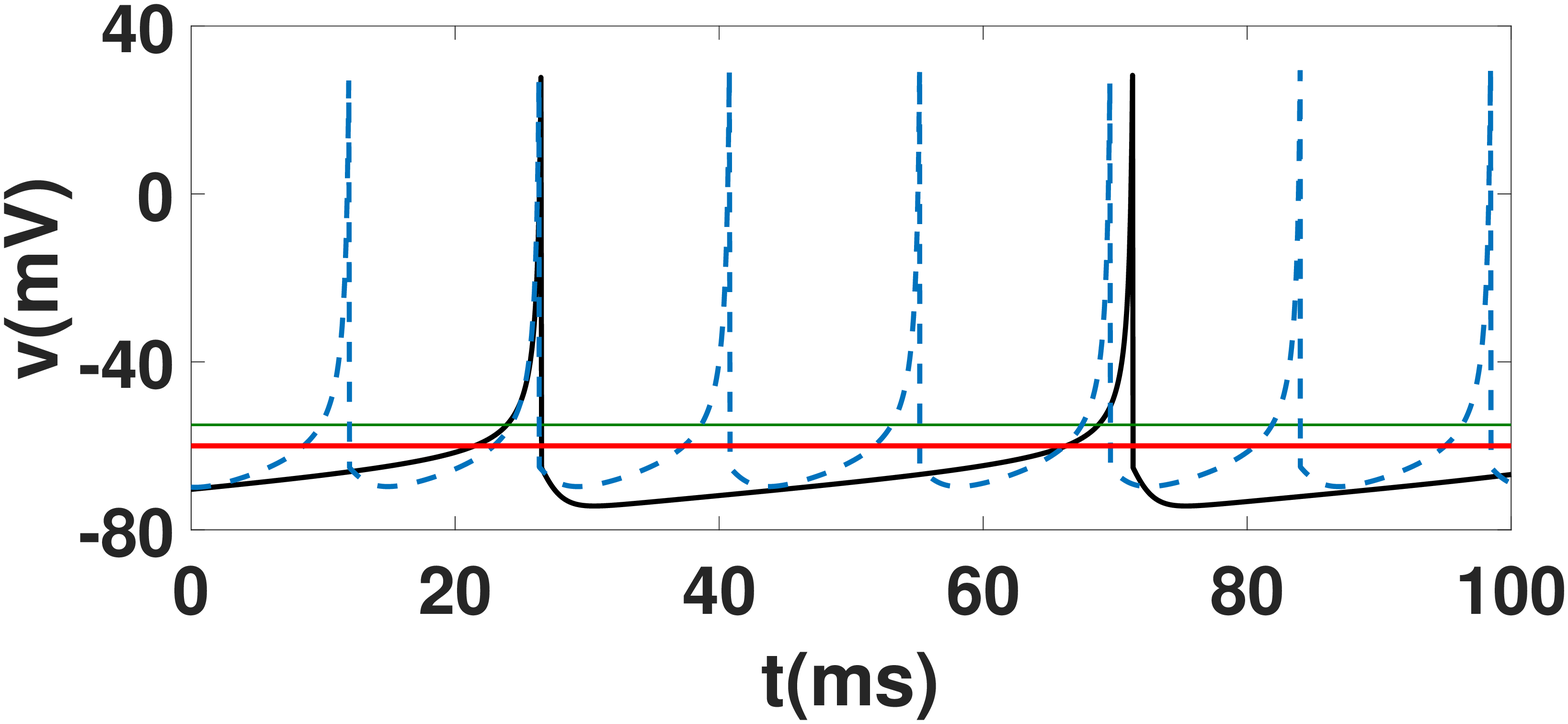}\label{fig7a}}
\subfigure[]{\includegraphics[width=0.48\textwidth,height=0.25\textwidth]{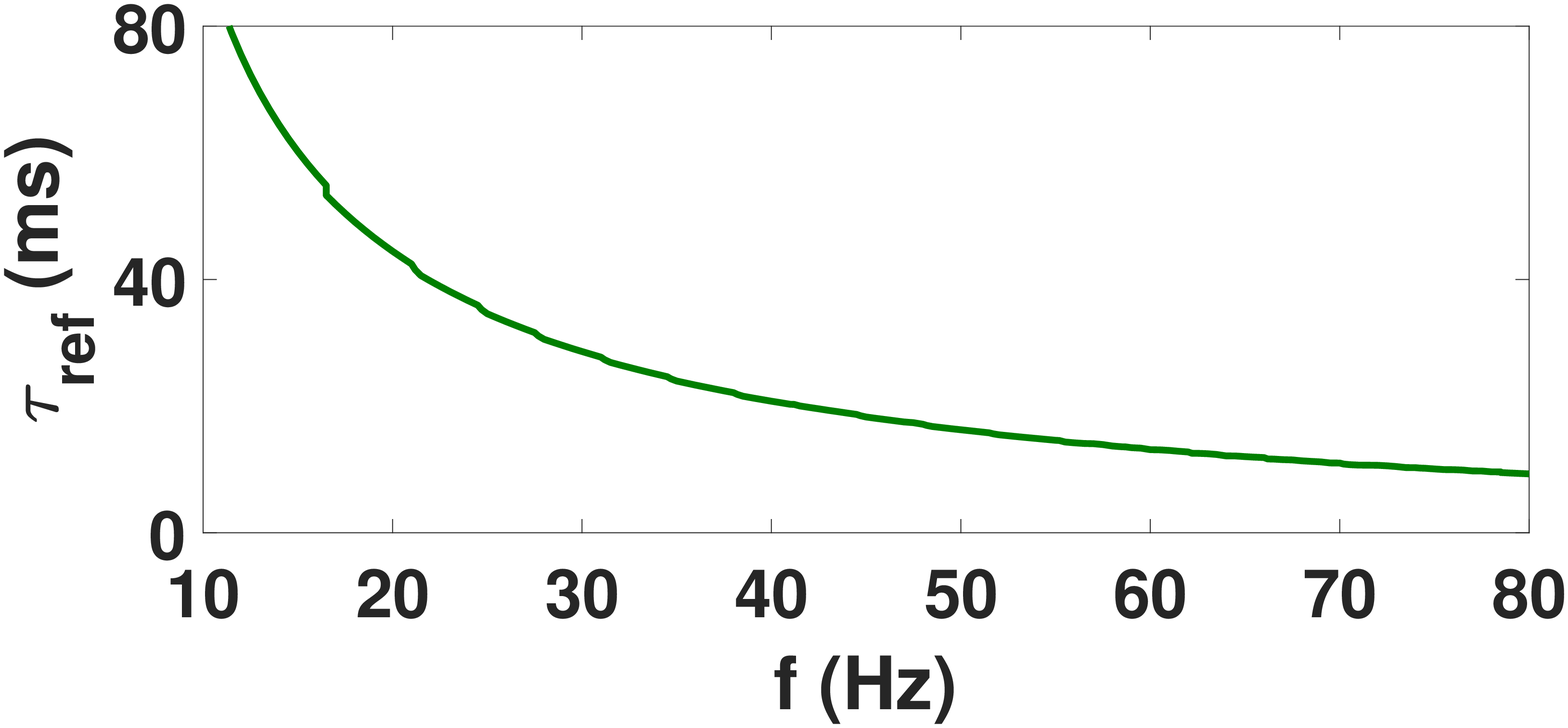}\label{fig7b}}
\end{center}
\caption{\small (a) Spike train of an Izhikevich neuron with
spiking frequency $22$ Hz (solid black curve) and with frequency
$70$ Hz (dashed blue curve). The solid green line indicate the
threshold of firing and solid red line is the rest potential. (b)
Dependence of refractory period $\tau_{ref}$ of Izhikevich neuron
on spiking frequency. $t=0$ indicates the beginning of stationary
state.} \label{fig7}
\end{figure}

In order to justify the frequency-dependent behavior of our neural
networks, we illustrate spike trains of an individual Izhikevich
neuron for two different values of firing frequencies $f=22$ Hz
and $f=70$ Hz in Fig.\ref{fig7a}. The horizontal solid green line
at -55 mV indicates the threshold for firing, while the solid red
line at -60 mV indicates the resting potential.  One can see that
the firing pattern of the two neurons are exactly the same, i.e.
the dynamics above threshold are identical.  However, the dynamics
below the resting potential is decidedly different, as the lower
frequency beta oscillation takes much longer to reach resting
potential. Note that the hyperpolarization is stronger in the beta
regime and the relative refractory period (time during which the
system remains below resting potential) is clearly longer. This
time scale $\tau_{ref}$ which renders the neuron to be relatively
unexcitable is an important factor.  In other words, while
changing the frequency of Izhikevich neurons does not change the
time scale of firing, it has a strong influence on the refractory
period.  This relative change of the time scales (firing vs.
refractory) can provide an explanation for why anti-phase
oscillations (and consequently explosive synchronization) occur in
low frequency regime but not in the high frequency regime.  In
fact, existence of anti-phase oscillations have been attributed to
separation of time scales in models of epidemic spreading
\cite{HS2003}. In Fig.\ref{fig7b}, we plot the refractory period
of Izhikevich neurons as function of frequency in the beta and
gamma regime. One sees that in the Izhikevich neuron, the
refractory period can become considerably long as one lowers the
frequency of oscillations.  While such a behavior may be an
artifact of the model, one can see that many other neuronal
dynamics models also exhibit similar behavior, i.e. a long time
associated with slow increase in potential at low frequencies.
Therefore, one might suspect that anti-phase oscillations and
explosive synchronization might be associated with other generic
neuronal models as well.

\section{discussion}
Synchronization transition in a network of oscillators has
attracted much attention in recent years. The Kuramoto model has
been used extensively in this regard with important implications
for neural networks.  However, it is a very crude approximation to
consider neurons as phase oscillators. In this work we have
studied synchronization transition in network models of
biologically plausible neurons. We used Izhikevich neurons in beta
frequency range coupled with electrical and chemical synapses on
various network structures. We found that stronger electrical
synapses are more conducive to synchronize than chemical synapses,
regardless of network structure. We also found that electrical
synapses can lead to anti-phase synchronization while no such
behavior was seen for chemical synapses. As far as network
structure was concerned, we found that the clustering coefficient,
and not the small-world effect, is the key topological factor that
determines the order of synchronization transition. When we
introduced short-cuts into the regular ring, no significant change
in the pattern of transition was observed. However, when random
networks with small clustering coefficient were considered,
synchronization patterns were significantly different.
Additionally, heterogeneity in network structure did not play an
important role as ER and SF results were identical. We note that
anti-phase synchronization leading to explosive synchronization
transition is a new and interesting mechanism which has not been
reported in the existing literature to the best of our knowledge.
The standard mechanisms reported in the literature for explosive
synchronization are associated with heterogeneity and disorder.
The reported results in this work becomes more interesting when we
note that explosive synchronization occurred in the beta band and
was not observed in the gamma band, i.e. it is frequency-dependent
and therefore of dynamical origin, as opposed to the more
widely-studied structural underpinnings. We note the fact that
beta and gamma rhythms have different synchronization patterns has
been reported before \cite{Kopell1999}. Such a frequency dependent
behavior in synchronization patterns seem important and deserves
further investigation \cite{KM2018}.

\section{Acknowledgements}
Support from Shiraz University research council is kindly
acknowledged. This work has been supported in part by a grant from
the Cognitive Sciences and Technologies Council. We would also
like to acknowledge useful conversations with Alireza Valizadeh.

\bibliographystyle{apsrev}

\end{document}